\documentclass[a4paper,11pt]{article}
\usepackage{jcappub}
\usepackage{amsmath,empheq}
\usepackage{hyperref}
\usepackage{graphicx}
\usepackage{tikz} 
\usepackage{float} 

\DeclareFontFamily{U}{BOONDOX-calo}{\skewchar\font=45 }
\DeclareFontShape{U}{BOONDOX-calo}{m}{n}{
	<-> s*[1.05] BOONDOX-r-calo}{}
\DeclareFontShape{U}{BOONDOX-calo}{b}{n}{
	<-> s*[1.05] BOONDOX-b-calo}{}
\DeclareMathAlphabet{\mathcalboondox}{U}{BOONDOX-calo}{m}{n}
\SetMathAlphabet{\mathcalboondox}{bold}{U}{BOONDOX-calo}{b}{n}
\DeclareMathAlphabet{\mathbcalboondox}{U}{BOONDOX-calo}{b}{n}

\ExplSyntaxOn
\NewDocumentEnvironment{sequation}{O{\small}b}
{
	\yufip_sequation:nnn {equation}{#1}{#2}
}{}
\NewDocumentEnvironment{sequation*}{O{\small}b}
{
	\yufip_sequation:nnn {equation*}{#1}{#2}
}{}
\cs_new_protected:Nn \yufip_sequation:nnn
{
	\begin{#1}
		\mbox{#2$\displaystyle#3$}
	\end{#1}
}
\ExplSyntaxOff

\title{Analytical Insights into Constant-Roll Condition: Extending the Paradigm to Non-Canonical Models}

\author[a]{S. Mohammad Ahmadi,}
\author[a]{Nahid Ahmadi}
\author[a,b,c]{and Mehdi Shokri}

\affiliation[a]{Department of Physics, University of Tehran, Karegar Ave North, Tehran 14395-547, Iran}
\affiliation[b]{School of Physics, Damghan University, P. O. Box 3671641167, Damghan, Iran}
\affiliation[c]{Canadian Quantum Research Center, 204-3002 32 Ave Vernon, BC V1T 2L7 Canada}

\emailAdd{mohammadahmadi@ut.ac.ir}
\emailAdd{nahmadi@ut.ac.ir}
\emailAdd{mehdishokriphysics@gmail.com}

\abstract{In this work, we explore the prospect of generalizing the constant-roll condition in canonical inflationary model to non-canonical models. To find a natural generalization, we focus on three manifestations of this condition and construct constant-roll models corresponding to each manifestation. These models are not equivalent but reduce to the familiar constant-roll model in canonical limit. To showcase the applicability of our generalized mechanism, we examine a specific class of non-canonical models, which can be viewed as extensions of k/G inflation. In these models sound speed is constant. We conduct a comparative study, and with an analytical examination of the model, specify instances when our constant-roll conditions yield dissimilar outcomes and when they exhibit analogies. We also apply our findings to scrutinize another kinetically driven inflationary model with varying sound speed. We demonstrate that each of our constant-roll conditions leads to a unique set of solutions. Afterward, we construct a four-stage constant-roll kinetically driven inflation that complies with CMB constraints, it sustains for a sufficiently long period of time, and finally gracefully exits. In this model the spectrum of curvature perturbations is enhanced in a brief phase of non-slow-roll inflationary evolution. Employing numerical methods, we analyse this scenario to elucidate how altering the constant-roll condition impacts the power spectrum and the model's dynamics.}

\begin{document}
\maketitle

\section{Introduction}\label{sec:introduction}

It is widely accepted that inflation provides the best explanation for the early universe \cite{starobinsky1980new, 10.1093/mnras/195.3.467, PhysRevD.23.347, Linde:1981mu, PhysRevLett.48.1437, LINDE1982389, LYTH19991, Martin_2016}. It proposes that a small epoch of accelerated expansion at very early universe can potentially smooth out inhomogeneities and anisotropies and reduce the curvature of space. In addition to explaining some of our fundamental questions, such as the flatness, horizon, and monopole problems, it suggests a mechanism for the generation of quantum perturbations in the early universe. There have been countless studies conducted on inflation, and observations of the cosmic microwave background (CMB) and of the large scale structure (LSS) of our universe have confirmed the predictions of inflation. In conventional inflationary models, a scalar field $\phi$ whose potential energy dominates the energy density of the universe is responsible for an accelerated expansion. The spectral index of canonical single field inflationary model is shown to be consistent with observational data in the slow-roll and ultra slow-roll (USR) regimes \cite{tsamis2004improved, kinney2005horizon, namjoo2013violation}. Nevertheless, the detailed dynamical characteristics of inflation is still obscure. Motivated by this, a one-parameter class of models (defined by the relation $\ddot{\phi}= \beta H \dot{\phi}$ or equivalently $\ddot{H} = 2 \beta H \dot{H}$, where $\beta$ is the free parameter of the model) was proposed in \cite{martin2013ultra} to unify simple scenarios like slow-roll and USR, and construct or maybe rule out many models of inflation. Later on, the same class of models was studied in more detail and referred to as \textit{constant-roll inflation} in \cite{motohashi2015inflation}. In the canonical constant-roll scenario, slow-roll behavior can be observed when $\beta \rightarrow 0$ and a constant USR potential at $\beta = -3$. Therefore, it can be considered a generalization of USR inflation.

The constant-roll scenario has attracted a great deal of attention in recent years. The novelty is that the equations of motion of the canonical constant-roll model can be solved analytically. Many instances of non-canonical constant-roll models can also be treated analytically. Its flexibility is another noteworthy feature of the model; changing the constant-roll parameter $\beta$ will allow one to switch between slow-roll and USR regimes. Different aspects of the canonical constant-roll model has been studied in \cite{martin2013ultra, motohashi2015inflation, cicciarella2018new, anguelova2018systematics, yi2018constant, morse2018large, pattison2018attractive, ghersi2019observational, boisseau2019inflationary, lin2019dynamical, firouzjahi2019stochastic, sadeghi2021anisotropic, oikonomou2022generalizing, stein2023simple, mohammadi2023constant, motohashi2023squeezed}. Furthermore, numerous studies have been conducted on non-minimal and non-canonical models under the constant-roll assumption \cite{motohashi2017f, nojiri2017constant, oikonomou2017reheating, odintsov2017unification, chervon2018method, awad2018constant, karam2018constant, gao2018constant, mohammadi2018tachyon, yi2018inflation, mohammadi2019constant, elizalde2019logarithmic, motohashi2019constant, odintsov2019constant, mohammadi2020beta, antoniadis2020constant, mohammadi2020constant, oikonomou2020reviving, shokri2021nonminimal, shokri2021generalized, shokri2022quintessential, shokri2022constantfer, shokri2022constant, wu2022baryogenesis, garnica2022constant, alhallak2022natural, herrera2023galilean, panda2023constant, el2023constant, bourakadi2023observational}. The constant-roll scenario may also be a candidate for the formation of primordial black holes, which has been the subject of considerable number of research \cite{motohashi2020constant, ozsoy2020slope,ng2021constant, wu2022cosmic, kristiano2022ruling, karam2205anatomy, davies2022non, tomberg2023stochastic, mishra2023primordial, ozsoy2023inflation}.

Although many studies have attempted to generalize the action of the constant-roll model, some authors focused on the generalization of the constant-roll condition \cite{motohashi2017f, boisseau2019inflationary, motohashi2019constant, oikonomou2022generalizing}. For example, authors in reference \cite{boisseau2019inflationary} add a new free parameter to the constant-roll condition by writing $\ddot{\phi}=\dot{\phi} (\beta H  + \alpha)$. Using this condition, the conventional constant-roll is restored with $\alpha = 0$, while a broader class of constant-roll potentials is identified for $\alpha \neq 0$.

The generalization of constant-roll condition is much more intriguing in non-canonical models. In a non-canonical model, unlike the canonical case, $\beta = -3$ does not always lead to USR. Therefore, to study the USR limit of the model an special form of generalized constant-roll condition is required. A good example of such a generalization can be found in \cite{motohashi2017f}. The purpose of their study was to investigate the $f(R)$ gravity under the assumption of a constant rate of roll.‌ For this purpose they defined the constant-roll condition as $\ddot{F} = \beta H_J \dot{F}$, where $F=\mathrm{d} f/ \mathrm{d} R$ and subscript $J$ denotes the Jordan frame. With this condition, USR and slow-roll regimes amount to $\beta\rightarrow -3$ and $\beta\rightarrow 0$ limits, respectively, and the model is considered to be a natural generalization of USR. The point is that such a generalization can be produced in many ways. One can choose one or another constant-roll condition based on simplicity or aesthetic elegance. In the study of non-canonical models, it is also possible to define an infinite number of distinct constant-roll conditions, all leading to the conventional constant-roll condition  $\ddot{\phi} = \beta H \dot{\phi}$ in the canonical limit, yet exhibit varied implications upon departure from the canonical framework. As a result of this freedom, constant-roll condition can be defined in accordance with the needs of the individual to study the desired physical concepts.

In this work, we use different manifestations of canonical constant-roll inflation in order to impose three types of constant-roll conditions on a general $P(X,\phi)$ action. Our first approach is based on the fact that the constant-roll condition is defined to ensure a constant rate of roll for the field. The second approach establishes constant-roll condition that leads to a constant $\epsilon_{2}$, and for our third constant-roll condition, we introduce the condition that corresponds to the USR as $\beta $ approaches $-3$. Next, we present and carefully examine a novel class of constant-roll with constant sound speed in order to demonstrate how a model can be affected by changing the constant-roll condition. While each constant-roll condition generally leads to distinct solutions and consequences, we demonstrate that specific limits can be identified where all three conditions converge toward a quasi-canonical phase. We use the term "quasi-canonical," in the sense that at these limits slow-roll parameters and consequently the perturbations of the model mimics the canonical model while the action generally differs from the canonical case. A good example in this class of models is k/G inflation \cite{lin2020primordial} in which sound speed is equal to one.

We then broaden our horizons by investigating another model with non-standard kinetic term \cite{armendariz1999k, chiba2000kinetically, armendariz2001essentials, chiba2002tracking, malquarti2003new, rendall2006dynamics} under the constant-roll assumption. We demonstrate that the model can naturally transition between two phases: a non-canonical constant-roll phase and a canonical phase. While the characteristics of the canonical phase remain independent of the specific constant-roll condition employed, the non-canonical phase and the transition dynamics between the two phases are highly dependent on the chosen constant-roll condition. To provide a more vivid demonstration of the constant-roll condition's impact on the model's dynamics, we introduce a four-stage constant $k$ scenario with the potential to give rise to primordial black holes. We discuss how our scenario complies with CMB constraints and facilitates a graceful exit from inflation. We conclude that altering the constant-roll condition can modify the resulting power spectrum, thereby significantly influence on the primordial black holes abundance.

This paper is arranged as follows. In Section \ref{sec: Canonical CR Model}, we review the canonical constant-roll model. In Section \ref{sec: Three constant-roll Conditions}, we introduce three distinct constant-roll conditions for non-canonical models. We then explore the applications of our generalized constant-roll mechanism by examining a class of non-canonical model in section \ref{sec: constant-roll and Propagation Inflation}. We scrutinize the interrelationships in the trio of constant-roll conditions, elucidating their analogies and distinctions. Section \ref{sec: Constant k Inflation} investigates another non-canonical model as an example of models that show different dynamical behavior in this comparative study. Next, we construct a multi-stage constant-roll scenario with a brief phase of non-slow-roll inflationary evolution capable of producing primordial black holes. Numerical calculations of the power spectrum and other dynamical parameters are presented to understand how different generalizations of the constant-roll condition can impact the model. We conclude in section \ref{sec: Summary and Conclusion}.

\section{Canonical Constant-Roll Inflation}\label{sec: Canonical CR Model}
In this section we present a review of the canonical constant-roll model \cite{martin2013ultra, motohashi2015inflation, anguelova2018systematics, morse2018large, lin2019dynamical}. We examine the constant-roll solutions and discuss the primordial scalar perturbations generated in these inflationary models. In next sections, when we discuss more general constant-roll cases, we will follow the same steps as here.

We consider the simplest inflationary scenario that contains a single scalar field with the standard kinetic term minimally coupled to gravity. The dynamics of this model is described by the well-known  Friedmann and Klein-Gordon equations,\footnote{Throughout the paper we work in Planck units where $c=G=\hbar=1$.}
\begin{align}
	& 3 H^{2}=\frac{\dot{\phi}^{2}}{2}+V(\phi), \label{Canonical CR - equation of motion 1} \\
	& \ddot{\phi}+3 H \dot{\phi}+V_{, \phi}=0, \label{Canonical CR - equation of motion 3}
\end{align}
where a dot stands for the derivative with respect to cosmic time. Furthermore, we have
\begin{equation}
	\dot{H}=-\frac{\dot{\phi}^{2}}{2}. \label{Canonical CR - equation of motion 2}
\end{equation}

To characterize the evolution of the background, slow-roll parameters are defined as follows:
\begin{equation}
	\epsilon_{1}  \equiv -\frac{\mathrm{d} \ln H}{\mathrm{~d} N} ,\quad \quad \epsilon_{j+1}\equiv\frac{\mathrm{d} \ln \epsilon_{j}}{\mathrm{~d} N},\label{Canonical CR - definition of SR parameters}
\end{equation}
where $j$ is a positive integer, and $N$ is the number of $e$-folds defined as $\mathrm{~d} N \equiv \mathrm{d} \ln a = H \mathrm{d} t$. Using equation \eqref{Canonical CR - equation of motion 2} and its derivative, it is easy to see that the second slow-roll parameter can be written as
\begin{equation}
	\epsilon_{2}  =2 \epsilon_{1}+\frac{\ddot{H}}{H \dot{H}}. \label{Canonic CR - 2nd SR parameter}
\end{equation}
Let us now consider a class of models in which the second term on the right-hand side of equation \eqref{Canonic CR - 2nd SR parameter} remains constant during the inflationary epoch
\begin{equation}
	\beta = \frac{\ddot{H}}{2 H \dot{H}}=\frac{\ddot{\phi}}{H \dot{\phi}}= \frac{\dot{V}}{2 H \dot{H}} - 3, \label{canonic CR - constant-roll condition}
\end{equation}
where $\beta$ is a real constant, and for the last two equity we have used equations \eqref{Canonical CR - equation of motion 3} and \eqref{Canonical CR - equation of motion 2}. Equation \eqref{canonic CR - constant-roll condition} is called the \textit{constant-roll condition} and $\beta$ is the \textit{constant-roll parameter}. From \eqref{canonic CR - constant-roll condition} we notice that $\beta$ constrains not only the rate of change of $\dot{H}$ and $\dot{\phi}$, but also the rate of change of potential (compared to that of $H^2$). Based on equations \eqref{Canonic CR - 2nd SR parameter} and \eqref{canonic CR - constant-roll condition}, and assuming that $\epsilon_{1}$ is small, $\epsilon_{2}$ can be approximated as $\epsilon_{2}\approx 2 \beta$. This class of models allows to set the parameter $\beta$ to zero to study the standard slow-roll scenario and go beyond to have extremely flat potential $V_{,\phi}=0$ for $\beta = -3$,
which is dubbed as \textit{ultra slow-roll inflation} (USR) \cite{tsamis2004improved, kinney2005horizon, namjoo2013violation}. In canonical models, the three purportedly distinct expressions for $\beta$ are actually 'the same' related by the system of equations \eqref{Canonical CR - equation of motion 1}-\eqref{Canonical CR - equation of motion 2}.

In this work, we often find it more convenient to work with number of $e$-folds $N$ rather than cosmic time $t$. Writing all time derivatives in equations \eqref{Canonical CR - equation of motion 1}-\eqref{Canonical CR - equation of motion 2}, and \eqref{canonic CR - constant-roll condition} in terms of $N$, we get
\begin{align}
	& 3 H^{2}=\frac{H^{2} \phi_{, N}^{2}}{2}+V, \label{Canonical CR - equation of motion N1} \\
	& H_{, N}=-\frac{H \phi_{, N}^{2}}{2}, \label{Canonical CR - equation of motion N2} \\
	& H^{2} \phi_{, N N}+H H_{, N} \phi_{, N}+3 H^{2} \phi_{, N}+V_{, \phi}=0, \label{Canonical CR - equation of motion N3}\\
	& H H_{, N N}+\left(H_{, N}\right)^{2}=2 \beta H H_{, N}. \label{Canonical CR - constant-roll condition N}
\end{align}
By integrating equation \eqref{Canonical CR - constant-roll condition N} for the Hubble parameter we find
\begin{equation}
	H^2=C_1 \left(C_2 e^{2 \beta N} + 1 \right), \label{Canonical CR - General solution of Hubble}
\end{equation}
where in general $C_1$ and $C_2$ are two arbitrary complex numbers. However, since the Hubble parameter is real, only real values for $C_1$ and $C_2$ are physically feasible. For simplicity, we rewrite these constants as $C_1 = \text{sgn}\left( C_1\right) M^2 $ and $C_2 = \text{sgn}\left( C_2\right) e^{2 \beta N_0}$, where $N_0$ is a real constant, $M$ is a positive real constant, and $\text{sgn}$ is the sign function. From this, we can deduce that the modulus of $C_2$
acts as a time shift. Therefore, without loss of generality, we can
set  $N_0 = 0$. Taking this into account, relation \eqref{Canonical CR - General solution of Hubble} can be rewritten as $H^2 = M^2 \text{sgn}\left( C_1\right) \left(\text{sgn}\left( C_2\right) e^{2 \beta N} +1 \right)$, which leaves us with the following choices:
\begin{align}
	H^2&=- M^2 \left(1+e^{2 \beta N}\right), \label{Canonical CR - hubble parameter new}\\
	H^2&=M^2 \left(1+e^{2 \beta N}\right), \label{Canonical CR - hubble parameter 1}\\
	H^2&=M^2 \left(1-e^{2 \beta N} \right), \label{Canonical CR - hubble parameter 2}\\
	H^2&=M^2 \left(e^{2 \beta N}-1\right). \label{Canonical CR - hubble parameter 3}
\end{align}
We first note that the right-hand side of solution \eqref{Canonical CR - hubble parameter new} is always negative, which leads to an imaginary Hubble parameter, and is therefore unacceptable. In contrast, solution \eqref{Canonical CR - hubble parameter 1} always results in a real Hubble parameter. For solutions \eqref{Canonical CR - hubble parameter 2} and \eqref{Canonical CR - hubble parameter 3}, the Hubble parameter is real only when we have $\beta N <0$ and $ \beta N >0$, respectively. Note that cases \eqref{Canonical CR - hubble parameter 2} and \eqref{Canonical CR - hubble parameter 3} are zero at $N=0$, which is the singular point of differential equation \eqref{Canonical CR - constant-roll condition N}. Therefore, our solutions are not viable in the vicinity of this point. As we will see later, despite the similarities between justifiable solutions, those describe valid inflationary dynamics for different ranges of $N$ and parameter $\beta$.

Having the Hubble parameter, we can derive the field and potential using equations \eqref{Canonical CR - equation of motion N1} and \eqref{Canonical CR - equation of motion N2}. For \eqref{Canonical CR - hubble parameter 1} case we get
\begin{align}
	&\phi(N)=\pm \sqrt{-\frac{2}{\beta}} \, \text{arcsinh} \left(e^{\beta N}\right)+\phi_{c}, \label{Canonical CR - first case - field} \\
	&V(\phi)=M^2 \left[3+ \left(\beta+3\right) \sinh ^2\left(\sqrt{-\frac{\beta}{2}} \left(\phi-\phi_{c}\right) \right)\right], \label{Canonical CR - first case - potential} \\
	&\ddot{a}(N)=a M^2 \left(1+(\beta+1) e^{2 \beta N }\right), \label{Canonical CR - first case - acceleration}
\end{align}
where $\phi_c$ is a constant of integration, and in \eqref{Canonical CR - first case - acceleration} we have used the relation $\mathrm{d}a/\mathrm{d} N = a$. This case corresponds to the second case studied in \cite{motohashi2015inflation}, as is evident when comparing the potential \eqref{Canonical CR - first case - potential} with equation (19) in their work. Let us now check whether this solution describes a viable inflationary period. To have real field, the constant-roll parameter $\beta$ should be negative. According to equation \eqref{Canonical CR - equation of motion N2}, for a real $\phi_{,N}$, $H_{,N}$ must be negative, which occurs when $\epsilon_{1}$ is positive,
and to have a positive acceleration during inflation, $\epsilon_{1}$ must be smaller than 1 (this point is easy to understand when one writes $\ddot{a}=a H^2 (1-\epsilon_{1})$). Lastly, using equations \eqref{Canonical CR - equation of motion 1}, \eqref{Canonical CR - equation of motion 2}, and \eqref{Canonical CR - definition of SR parameters}, potential can be written as $V/H^2= 3- \epsilon_{1}$; so the potential is positive unless $\epsilon_1$ exceeds $3$. But this never happens with $0<\epsilon_{1} < 1$.

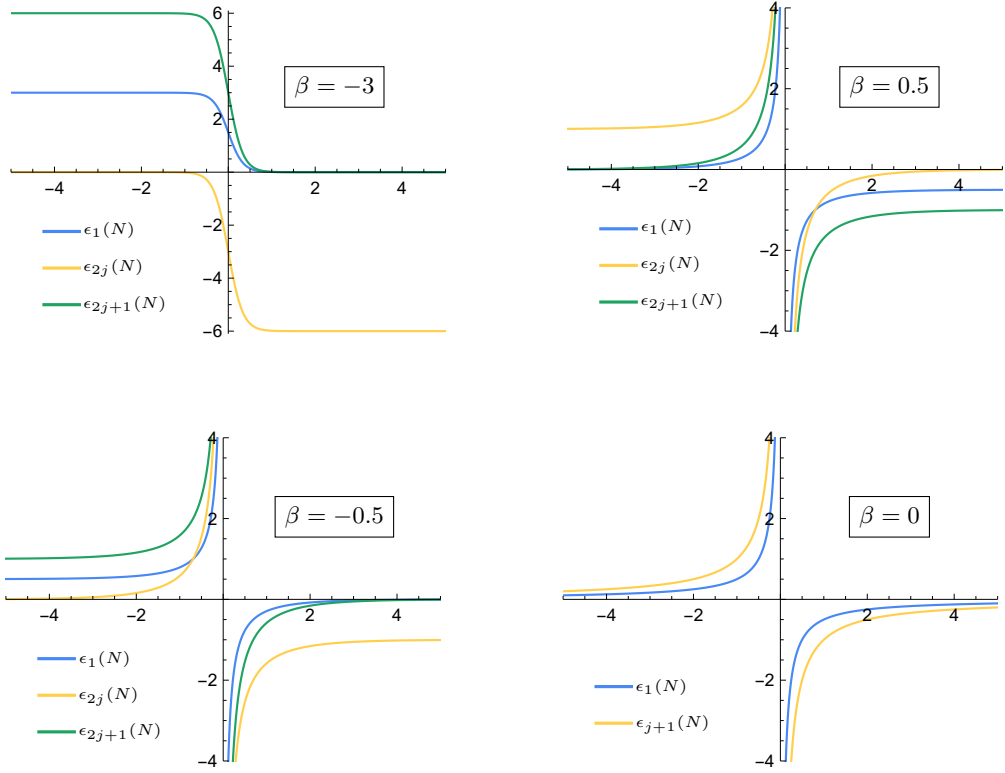
\begin{figure}
	\centering
	\begin{tikzpicture}
		\node[align=left] (img) at (0,5.7)
		{\quad\,\,\includegraphics{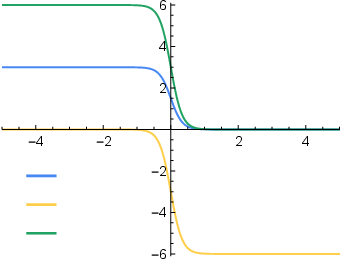}\quad\quad\quad\quad\includegraphics{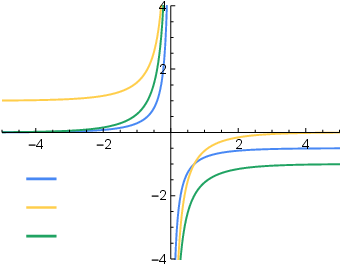}\quad};
		\node[align=left] (img) at (0,0)
		{\quad\includegraphics{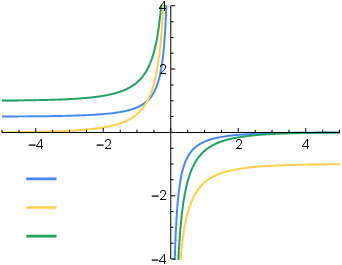}\quad\quad\quad\quad\includegraphics{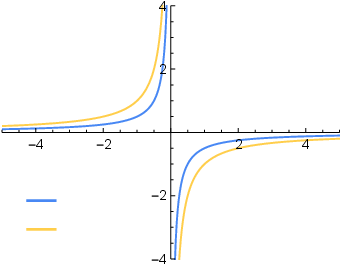}\quad}; //
		
		\node at (-5.16,+4.90) {{\tiny $\epsilon_{1}(N)$}};
		\node at (-5.11,+4.40) {{\tiny $\epsilon_{2j}(N)$}};
		\node at (-4.96,+3.91) {{\tiny $\epsilon_{2j+1}(N)$}};
		\node [draw] at (-2.2,+6.8) {{{\footnotesize $\beta=-3$}}};
		
		\node at (+2.17,+4.94) {{\tiny $\epsilon_{1}(N)$}};
		\node at (+2.22,+4.44) {{\tiny $\epsilon_{2j}(N)$}};
		\node at (+2.37,+3.95) {{\tiny $\epsilon_{2j+1}(N)$}};
		\node [draw] at (+5.1,+6.8) {{{\footnotesize $\beta=0.5$}}};
		
		\node at (-5.22,-0.79) {{\tiny $\epsilon_{1}(N)$}};
		\node at (-5.17,-1.28) {{\tiny $\epsilon_{2j}(N)$}};
		\node at (-5.02,-1.76) {{\tiny $\epsilon_{2j+1}(N)$}};
		\node [draw] at (-2.2,+1.1) {{{\footnotesize $\beta=-0.5$}}};
		
		\node at (+2.10,-1.16) {{\tiny $\epsilon_{1}(N)$}};
		\node at (+2.25,-1.65) {{\tiny $\epsilon_{j+1}(N)$}};
		\node [draw] at (+5.1,+1.1) {{{\footnotesize $\beta=0$}}};
		
	\end{tikzpicture}
	\caption{Slow-roll parameters for the canonical constant-roll model. Parameters \eqref{Canonical CR - SR parameters for solution 1}, \eqref{Canonical CR - SR parameters for solution 2}, \eqref{Canonical CR - SR parameters for solution 3}, and \eqref{Canonical CR - SR parameters for solution 4} are plotted in the top left, top right, bottom left, and bottom right panels, respectively.}
	\label{fig:Canonic CR - SR parameters}
\end{figure}

After substituting the Hubble parameter \eqref{Canonical CR - hubble parameter 1} into equations \eqref{Canonical CR - definition of SR parameters}, the slow-roll parameters can be derived as follows:
\begin{align}
	\epsilon_1 (N)  =\frac{-\beta}{1+e^{-2 \beta N}}, 
	\quad\quad
	\epsilon_{2 j}(N)  =\frac{2 \beta}{1+ e^{2 \beta N}},
	\quad\quad
	\epsilon_{2 j+1}(N)  =2 \epsilon_{1} (N). \label{Canonical CR - SR parameters for solution 1}
\end{align}
Slow-roll parameters \eqref{Canonical CR - SR parameters for solution 1} are illustrated in the top left panel of Figure \ref{fig:Canonic CR - SR parameters} for $\beta = -3$. In large negative $N$, the first slow-roll parameter can be approximated as $\epsilon_{1} \approx -\beta$, and in large positive $N$ it tends to zero. That means for $-1<\beta<0$ condition $0<\epsilon_{1}<1$ is always satisfied. However, for $\beta < -1 $, $\epsilon_{1}$ is smaller than one only when $2 \beta N < \ln \left(-1/(\beta+1)\right)$. Moreover, it is clear that $\beta$ cannot be positive as it leads to negative $\epsilon_{1}$. Consequently, inflation conditions are met for $-1\le\beta<0$ for all $N$ values, or $\beta < -1$ at $N > \ln \left(-1/(\beta+1)\right) / (2 \beta)$.

Similarly we can discuss the acceptable ranges for solutions \eqref{Canonical CR - hubble parameter 2} and \eqref{Canonical CR - hubble parameter 3}. For \eqref{Canonical CR - hubble parameter 2} case we can find
\begin{align}
	&\phi(N)=\pm \sqrt{\frac{2}{\beta}} \, \text{arccos}\left(e^{\beta N}\right)+\phi_{c}, \label{Canonical CR - Solution 21} \\
	&V(\phi)=M^2 \left[3- \left(\beta+3\right) \cos ^2\left(\sqrt{\frac{\beta}{2}} \left(\phi-\phi_{c}\right) \right)\right], \label{Canonical CR - Solution 22} \\
	&\ddot{a}(N)=a M^2 \left(1-(\beta+1) e^{2 \beta N }\right). \label{Canonical CR - Solution 23}
\end{align}
And the slow-roll parameters are
\begin{equation}
	\epsilon_1 (N)  =\frac{-\beta}{1-e^{-2 \beta N}}, 
	\quad\quad
	\epsilon_{2 j}(N)  =\frac{2 \beta}{1 - e^{2 \beta N}},
	\quad\quad
	\epsilon_{2 j+1}(N)  =2 \epsilon_{1} (N). \label{Canonical CR - SR parameters for solution 2}
\end{equation}
Considering potential \eqref{Canonical CR - Solution 22}, we can readily observe that this case corresponds to the third case studied in the original work \cite{motohashi2015inflation} (see equation (23) therein). Figure \ref{fig:Canonic CR - SR parameters} (top right panel) shows parameters \eqref{Canonical CR - SR parameters for solution 2} for $\beta=0.5$. In this case, for the allowed values of $\beta N$, $\epsilon_{1}$ is positive only if $N<0$ and $\beta>0$. At $N = 0 $, however, $\epsilon_{1}$ is singular and moves to infinity. Therefore, for the condition $\epsilon_{1}<1$ to be satisfied, $2 \beta N$ must remain smaller than $\ln \left(1/(\beta+1)\right)$. For \eqref{Canonical CR - hubble parameter 3} case we get
\begin{align}
	&\phi(N)=\pm \sqrt{-\frac{2}{\beta}} \, \text{arccosh}\left(e^{\beta N}\right)+\phi_{c}, \label{Canonical CR - Solution 31} \\
	&V(\phi)=M^2 \left[ \left(\beta+3\right) \cosh ^2\left(\sqrt{-\frac{\beta}{2}} \left(\phi-\phi_{c}\right) \right)-3\right], \label{Canonical CR - Solution 32} \\
	&\ddot{a}(N)=a M^2 \left( (\beta+1) e^{2 \beta N } -1\right), \label{Canonical CR - Solution 33}
\end{align}
and for slow-roll parameters we have
\begin{equation}
	\epsilon_1 (N)  =\frac{-\beta}{1-e^{-2 \beta N}}, 
	\quad\quad
	\epsilon_{2 j}(N)  =\frac{2 \beta}{1 - e^{2 \beta N}},
	\quad\quad
	\epsilon_{2 j+1}(N)  =2 \epsilon_{1} (N). \label{Canonical CR - SR parameters for solution 3}
\end{equation}
Inspection of the potential \eqref{Canonical CR - Solution 32} reveals that this final case coincides with the special case analyzed in reference \cite{anguelova2018systematics}. Parameters \eqref{Canonical CR - SR parameters for solution 3} are plotted in Figure \ref{fig:Canonic CR - SR parameters} (bottom left panel) for $\beta=-0.5$. Note that for the allowed values of $\beta N$, $\epsilon_{1}$ is positive only if $N <0 $ and $\beta <0$. At large negative $N$ we can approximate $\epsilon_{1}$ to $-\beta$. In this limit condition $0<\epsilon_{1} < 1$ leads to $-1<\beta<0$. On the other hand, when $N \rightarrow 0$, $\epsilon_{1}$ moves to infinity, and condition $2 \beta N > \ln \left(1/(\beta+1)\right)$ is required to keep $\epsilon_{1}$ smaller than $1$.

It is noteworthy that $\beta=-3$ is not allowed in potentials \eqref{Canonical CR - Solution 22} and \eqref{Canonical CR - Solution 32}, whereas it is allowed in potential \eqref{Canonical CR - first case - potential}. It follows that only potential \eqref{Canonical CR - first case - potential} can represent a case of USR inflation.

Heretofore, we analysed the general solution of equation \eqref{Canonical CR - constant-roll condition N}. However, our solutions are singular at $\beta=0$ (see equations \eqref{Canonical CR - first case - field}, \eqref{Canonical CR - Solution 21}, and \eqref{Canonical CR - Solution 31}).
To examine the model at this parameter value, we integrate the constant-roll condition \eqref{Canonical CR - constant-roll condition N} again, this time after setting $\beta$ to zero. We will find
\begin{equation}
	H^2= C \left(N - N_0\right),
\end{equation}
where $C$ and $N_0$ are in general two complex numbers. Taking into account that the Hubble parameter is real, we are left with the following choices:
\begin{align}
	&H^2=- M^2 N, \label{Canonical CR - hubble parameter 0}\\
	&H^2=+ M^2 N, \label{Canonical CR - hubble parameter zero}
\end{align}
where $M^2 = |C|$, and without losing generality we have assumed that inflation starts at $N= N_{\text{ini}}$ and have set $N_0$ to zero. We first note that solution \eqref{Canonical CR - hubble parameter zero} is not acceptable as it leads to a positive $H_{,N}$ and an imaginary $\phi_{, N}$. For solution \eqref{Canonical CR - hubble parameter 0} the field, potential, and acceleration are as follows:
\begin{align}
	&\phi(N)=\pm 2 \sqrt{-N}+\phi_{c}, \label{Canonical CR - Solution 41} \\
	&V(\phi)=\frac{1}{4} M^2 \left[ 3 \left(\phi-\phi_{c}\right)^2 - 2 \right], \label{Canonical CR - Solution 42} \\
	&\ddot{a}(N)=-\frac{1}{2} a M^2 \left( 1 + 2 N \right),  \label{Canonical CR - Solution 43}
\end{align}
and for the slow-roll parameters we get
\begin{equation}
	\epsilon_{1}(N)=-\frac{1}{2 N},
	\quad\quad
	\epsilon_{j+1}(N)=2 \epsilon_{1}(N). \label{Canonical CR - SR parameters for solution 4}
\end{equation}
The slow-roll parameters \eqref{Canonical CR - SR parameters for solution 4} are plotted in the bottom right panel of Figure \ref{fig:Canonic CR - SR parameters}. In this case inflation conditions are satisfied for $N< -1/2$. With this last solution, we have completely discussed the dynamics of the canonical constant-roll model.

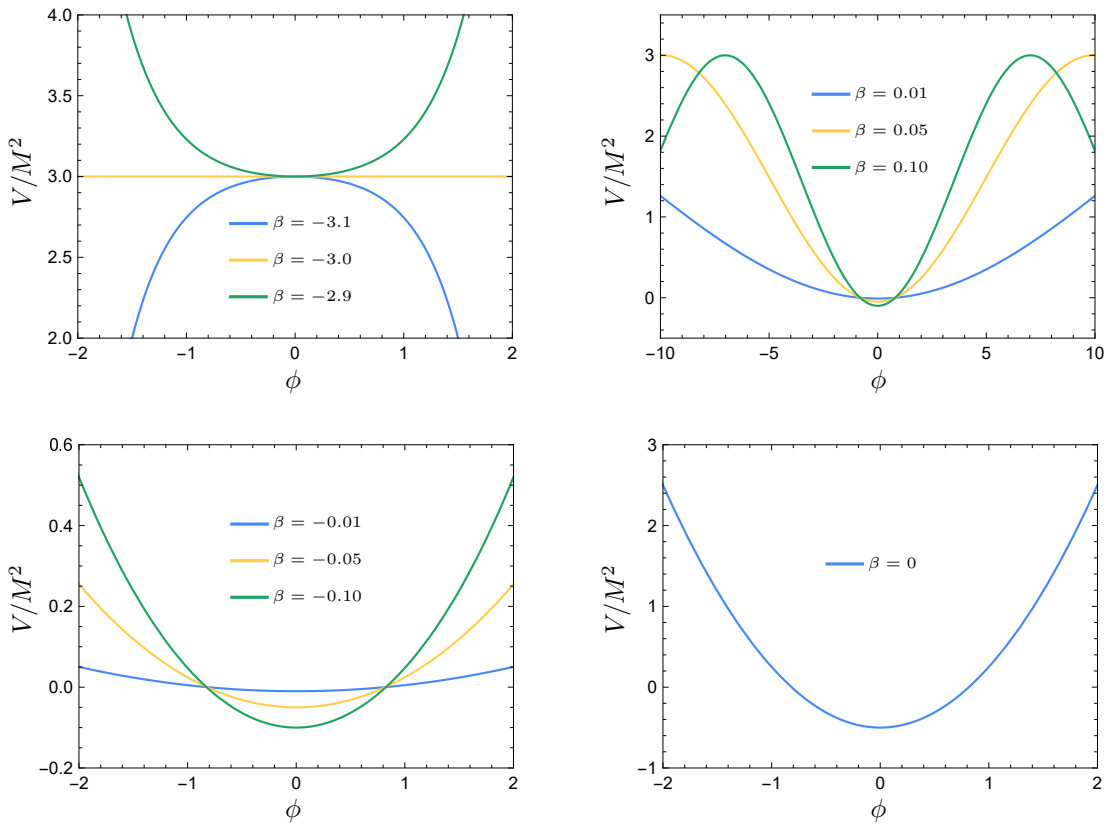
\begin{figure}
	\centering
	\begin{tikzpicture}
		\node[align=left] (img) at (0,5.7)
		{\quad\,\,\includegraphics{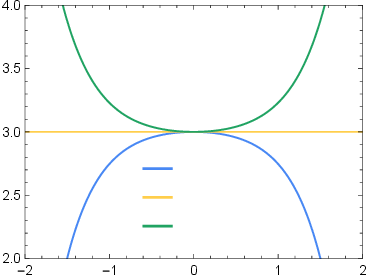}\quad\quad\quad\quad\,\,\includegraphics{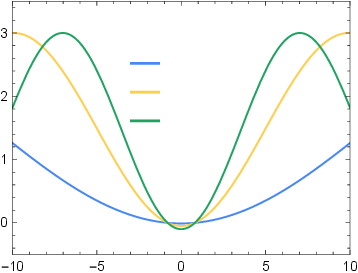}\quad};
		\node[align=left] (img) at (0,0)
		{\quad\includegraphics{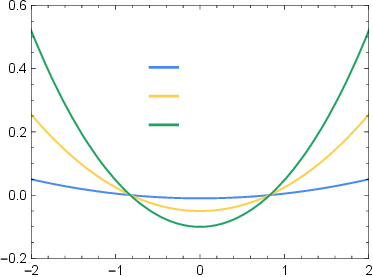}\quad\quad\quad\quad\includegraphics{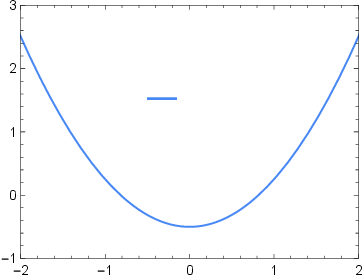}\quad}; //
		
		\node at (-7.2,0.21) {{\small \rotatebox{90}{$V/ M^2$}}};
		\node at (+0.6,0.21) {{\small \rotatebox{90}{$V/ M^2$}}};
		\node at (-7.2,5.88) {{\small \rotatebox{90}{$V/ M^2$}}};
		\node at (+0.6,5.88) {{\small \rotatebox{90}{$V/ M^2$}}};
		
		\node at (-3.65,-2.6) {{\small $\phi$}};
		\node at (+4.05,-2.6) {{\small $\phi$}};
		\node at (-3.65,+3.1) {{\small $\phi$}};
		\node at (+4.05,+3.1) {{\small $\phi$}};
		
		\node at (-3.40,+5.19) {{\tiny $\beta=-3.1$}};
		\node at (-3.40,+4.71) {{\tiny $\beta=-3.0$}};
		\node at (-3.40,+4.21) {{\tiny $\beta=-2.9$}};
		
		\node at (+4.22,+6.90) {{\tiny $\beta=0.01$}};
		\node at (+4.22,+6.42) {{\tiny $\beta=0.05$}};
		\node at (+4.22,+5.93) {{\tiny $\beta=0.10$}};
		
		\node at (-3.33,+1.21) {{\tiny $\beta=-0.01$}};
		\node at (-3.33,+0.72) {{\tiny $\beta=-0.05$}};
		\node at (-3.33,+0.23) {{\tiny $\beta=-0.10$}};
		
		\node at (+4.23,+0.67) {{\tiny $\beta=0$}};
	\end{tikzpicture}
	\caption{The class of canonical constant-roll potentials. Potentials \eqref{Canonical CR - first case - potential}, \eqref{Canonical CR - Solution 22}, \eqref{Canonical CR - Solution 32}, and \eqref{Canonical CR - Solution 42} are plotted in the top left, top right, bottom left, and bottom right panels respectively. The parameter $\phi_{c}$ is set to zero in all plots.}
	\label{fig:Canonic CR - V(N)}
\end{figure}
Figure \ref{fig:Canonic CR - V(N)} shows all constant-roll potentials \eqref{Canonical CR - first case - potential}, \eqref{Canonical CR - Solution 22}, \eqref{Canonical CR - Solution 32}, and \eqref{Canonical CR - Solution 42}. As can be seen, the potentials are extremely flat for $\beta \rightarrow -3$ (USR limit) and $\beta \rightarrow 0$ (slow-roll limit). But when $\beta$ is exactly zero, the potential has a parabolic shape. It is true that in the case of the slow-roll, as in the case of USR, the potential should be very flat to ensure a slow rate of roll and a minimum of 40-60 $e$-folds of inflation. But if one defines USR as a model with a flat potential, then $\beta \rightarrow 0$ will be an USR case as well, and there is no clear distinction between USR and slow-roll models. Therefore, in order to avoid ambiguities, we define USR limit with condition $V_{,\phi} \ll \{ H \dot{\phi},\ddot{\phi}\}$, and slow-roll limit with $\{\epsilon_{1},\epsilon_{2}, \text{...}\} \ll 1$ (which in the canonical case leads to $\ddot{\phi} \ll H \dot{\phi} \ll H^2$). 
By expanding the potential and field for solutions \eqref{Canonical CR - hubble parameter 1}, \eqref{Canonical CR - hubble parameter 2}, and \eqref{Canonical CR - hubble parameter 3} around 0 and $-3$, we find that when $\beta \rightarrow 0$ we have $\ddot{\phi} / V_{,\phi} \propto \beta$ $V_{,\phi}/\ddot{\phi} \propto 1/\beta$, while when $\beta \rightarrow -3$ we have $V_{,\phi}/\ddot{\phi} \propto \beta+3$ and $V_{,\phi}/\dot{\phi} \propto \beta+3$. Thus, $\beta \rightarrow 0$ does not represent an USR case but $\beta \rightarrow -3$ does. Solution \eqref{Canonical CR - hubble parameter 0} also is not an USR case as it leads to $\ddot{\phi}=0$ and condition $V_{,\phi} \ll \ddot{\phi}$ cannot be satisfied.

It could also be argued that the name "ultra slow-roll" is inappropriate. Because USR begins with a period of fast roll inflation. Nevertheless, we do not obsess over nomenclature in this work and respect the names that have long been used. Additionally, it should be noted that the model described by equations \eqref{Canonical CR - first case - field}-\eqref{Canonical CR - first case - acceleration} is an USR model for parameter value $\beta = -3$, but the reverse is not always true.
A similar argument can be made for the slow-roll limit $\beta \approx 0$.  In the slow-roll approximation $\ddot{\phi}/H \dot{\phi}$ is very small and not necessarily constant.

Our next step is to evaluate the spectral index and the non-Gaussianity parameter in order to examine the behavior of perturbations. In solution $\eqref{Canonical CR - hubble parameter 1}$ at large positive $N$ and solution $\eqref{Canonical CR - hubble parameter 2}$ at large negative $N$, $\epsilon_{1}$ and $\epsilon_{3}$ are small and $\epsilon_{2}$ is roughly $2 \beta$. Therefore, the spectral index can be approximated as \cite{namjoo2013violation, martin2013ultra, motohashi2015inflation, motohashi2017constant} (see Appendix \ref{Appendix: Power Specturm})
\begin{equation}
	n_{\mathrm{s}}-1= \begin{cases}2(\beta+3), & \beta < -3 / 2 \\ -2 \beta, &\beta \ge -3 / 2 \,\, \text{and} \,\,  \beta \neq 0 \end{cases}. \label{Canonical CR - spectral index}
\end{equation}
The 2018 Planck data \cite{akrami2020planck} constrain the spectral index as $n_s=0.9649 \pm 0.0042$.  Accordingly, the only feasible values for the constant-roll parameter are $\beta \approx 0.018$ and $\beta \approx -3.018$. It is imperative to note that for deriving \eqref{Canonical CR - spectral index} the power is evaluated at horizon exit. However, as we will show later, in the case $\beta < -3/2$ the modes continue to grow on the supperhorizon scales and the effect of growing modes should also be taken into account in the evaluation of the power spectrum. Consequently, for $\beta \approx -3.018$, result \eqref{Canonical CR - spectral index} is invalid, and $\beta \approx 0.018$ is the only compatible point.

In solutions \eqref{Canonical CR - hubble parameter 1} and \eqref{Canonical CR - hubble parameter 3} if inflation starts at large negative $N$ we have $\epsilon_{1} \approx -\beta$ and $\epsilon_{3} \approx -2 \beta$ while $\epsilon_{2}$ is small. So for spectral index we get \cite{noller2011non}
\begin{equation}
	n_{\mathrm{s}}-1=\frac{2\beta}{\beta+1}. \label{Canonical CR - spectral index 2}
\end{equation}
In this case, the constant-roll parameter is constrained to $\beta \approx -0.017$, based on Planck data. Finally, for solution \eqref{Canonical CR - hubble parameter 0}, since all slow-roll parameters are small, we will have the following spectral index:
\begin{equation}
	n_{\mathrm{s}}-1= -4 \epsilon_{1}. \label{Canonical CR - spectral index 3}
\end{equation}
To agree with the Planck data, $\epsilon_{1}$ should be in the range of $0.0077$ to $0.0098$, which is between $N=-65$ and $-51$. That means this solution cannot describe a complete inflationary epoch and should be accompanied with some stable phases. As far as we know, results \eqref{Canonical CR - spectral index 2} and \eqref{Canonical CR - spectral index 3} are new.

\begin{figure}
	\centering
	\begin{tikzpicture}
		\node[align=left] (img) at (0,2.5)
		{\quad\includegraphics[scale=0.6]{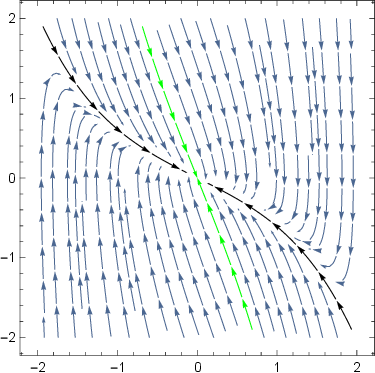}\quad\quad\quad\includegraphics[scale=0.6]{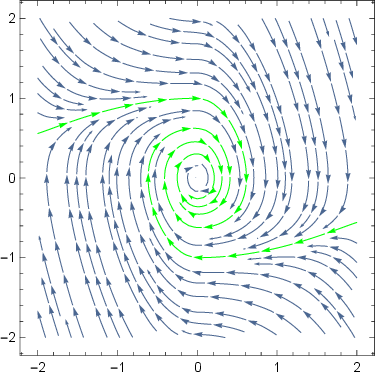}\quad\quad\quad\includegraphics[scale=0.6]{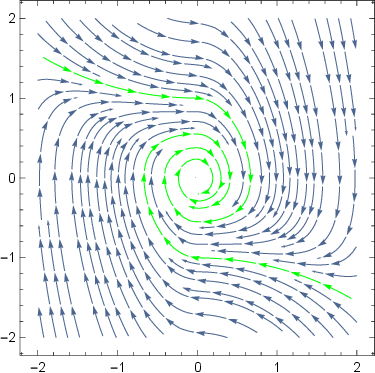}\quad}; //
		
		\node at (-7.10,2.64) {{\footnotesize $\dot{\phi}$}};
		\node at (-2.14,2.64) {{\footnotesize $\dot{\phi}$}};
		\node at (+2.83,2.64) {{\footnotesize $\dot{\phi}$}};
		
		\node at (-4.90,0.3) {{\footnotesize $\phi$}};
		\node at (+0.06,0.3) {{\footnotesize $\phi$}};
		\node at (+5.03,0.3) {{\footnotesize $\phi$}};
		
	\end{tikzpicture}
	\caption{Left, middle, and right panels show phase portrait diagrams for canonical constant-roll solutions \eqref{Canonical CR - first case - field}(for $\beta = -2.5$), \eqref{Canonical CR - Solution 21}(for $\beta = 0.5$), and \eqref{Canonical CR - Solution 31} (for $\beta = -0.5$). Green lines are unperturbed constant-roll trajectories. The black line in the left panel shows the corresponding dual constant-roll trajectory, $\beta \approx -0.5$.}
	\label{fig:Canonic CR - phase space diagrams}
\end{figure}
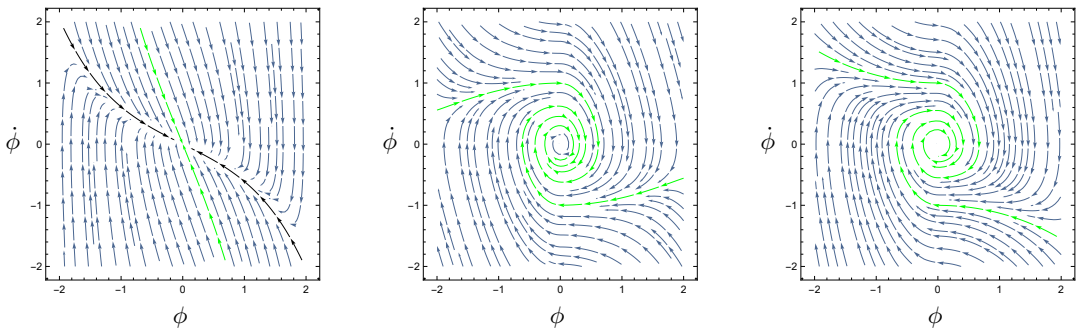

To determine whether our solutions are attractor and stable, we plot the phase space diagrams. For the first solution \eqref{Canonical CR - first case - field}, phase space is illustrated Figure \ref{fig:Canonic CR - phase space diagrams} (left panel). This picture clearly shows that the constant-roll trajectory for $\beta = -2.5$ (green line) is not an attractor, but its corresponding dual constant-roll solution is. Dual solutions are constant-roll solutions for $\tilde{\beta} \approx -3 - \beta$. References \cite{morse2018large, lin2019dynamical} provide more information about the duality of constant-roll solutions.

In Figure \ref{fig:Canonic CR - phase space diagrams}, the middle and right panels illustrate phase space diagrams for solutions \eqref{Canonical CR - Solution 21} and \eqref{Canonical CR - Solution 31}. As can be seen, these solutions are attractor and stable.

For solution \eqref{Canonical CR - Solution 41}, $\beta$ is taken to be exactly zero. In this case, a slight change in initial velocity will result in switching to another constant-roll solution and never returning to $\beta=0$ again. As a result, this solution is also unstable. In general, all $\beta < -3/2$ solutions are unstable, while all $\beta \ge -3/2$ solutions, except for $\beta=0$, are stable.

This work does not focus on studying non-Gaussianity. However, the non-Gaussian properties of solution \eqref{Canonical CR - hubble parameter 1} are particularly interesting and worth discussion. In this case, the non-Gaussianity parameter can be derived as follows \cite{namjoo2013violation, martin2013ultra} (see Appendix \ref{Appendix: Power Specturm}):
\begin{equation}
	f_{N L}^{r e}= \begin{cases}-\frac{5}{2}( \beta+2)=\frac{5}{4}\left(3-n_{s}\right), & \beta < -3 / 2 \\ \frac{5}{6} \beta=\frac{5}{12}\left(1-n_{s}\right), & \beta \ge -3 / 2 \,\, \text{and} \,\,  \beta \neq 0\end{cases}. \label{Canonical CR - non-Gaussianity parameter}
\end{equation}
Parameter \eqref{Canonical CR - non-Gaussianity parameter} is of order one for $\beta \approx -3.018$ and contributes the most to non-Gaussianity. However, when $\beta \approx 0.018$, this parameter is of order one for the slow-roll parameter and its contribution to non-Gaussianity is comparable to some terms in the action. Additionally, Maldacena's consistency relation \cite{maldacena2003non} is satisfied for $\beta \approx 0.018$, but violated for $\beta \approx -3.018$.

A look at the evolution of curvature perturbations on supper Hubble scales ($k \ll a H$) can help explain the sudden change of behavior in equations \eqref{Canonical CR - spectral index} and \eqref{Canonical CR - non-Gaussianity parameter} at $\beta = -3/2$. In de Sitter limit, the superhorizon curvature perturbation $\mathcal{R}$ is given by (see equation \eqref{Appendix power - supper hubble evolution} for $c_s = 1$)
\begin{equation}
	\mathcal{R}=C_1+C_2 \int \frac{\mathrm{~d} \tau}{a^2 \epsilon_1}=C_1+\frac{C_2}{a_0^3 H} \int \frac{\mathrm{~d} N}{e^{3N} \epsilon_1}, \label{Canonical CR - super hubble evolution of carvature perturbation}
\end{equation}
where $C_1$ and $C_1$ are two arbitrary constants and $a_0=a(N=0)$.
As can be seen in equation \eqref{Canonical CR - super hubble evolution of carvature perturbation}, the curvature perturbation has a constant part and an integral part.  If $\epsilon_{1}$ is decaying slower than $e^{-3N}$, then the mode function will converge to a constant at super Hubble scales. In contrast, when $\epsilon_{1}$ decays faster than $e^{-3N}$, the integral part diverges and the mode function continues to evolve at large scales. Consistency relation requires that the long modes are well frozen out at large scales, which is not the case here. Therefore, Maldacena's methodology for proving non-Gaussianity is small does not apply here, and consistency relation cannot be satisfied for fast decaying $\epsilon_{1}$. In equation \eqref{Canonical CR - SR parameters for solution 1}, it is evident that for $\beta = - 3/2$, the slow-roll parameter decays as $e^{-3N}$. This completely explains why  equations \eqref{Canonical CR - spectral index} and \eqref{Canonical CR - non-Gaussianity parameter} suddenly change their behavior at $\beta = - 3/2$.

\section{Constant-Roll Condition in Non-Canonical Models} \label{sec: Three constant-roll Conditions}
In the previous section, the constant-roll condition was defined by three equivalent expressions given in \eqref{canonic CR - constant-roll condition}. In this section, we consider each expression as a distinct constant-roll condition and impose it on a general $P(X,\phi)$ action to show that the nature of the model and its consequences is sensitive to the expression we use. Furthermore, the equivalence is restricted to some models, like single field canonical case, and the result would be different for either expression, in other inflationary models. This \textit{comparative} study shows how different characteristics of a constant-roll model are contingent upon the constancy of each expression. 

We consider a general action as follows:
\begin{equation}
	S=\frac{1}{2} \int d^{4} x \sqrt{-g}[R+2 P(X, \phi)], \label{General CR: general P(X) action}
\end{equation}
where $P(X,\phi)$ is an arbitrary function of $X$ and $\phi$, and $X=-\frac{1}{2} g^{\mu \nu} \partial_\mu \phi \partial_\nu \phi$. In FLRW background with $(-,+,+,+)$ signature, $X=\dot{\phi}^2 / 2$ corresponds to the canonical kinetic term. The Friedman equations for this action are
\begin{align}
	& 3 H^{2}=2XP_{,X}-P, \label{General CR - P(X) equation of motion 1} \\
	& \dot{H}+X P_{, X}=0.  \label{General CR - P(X) equation of motion 2}
\end{align}
The time evolution of the homogenous mode of the field $\phi(t)$ is governed by the Klein-Gordon equation
\begin{equation}
	\ddot{\phi}+3 c_{s}^{2} H \dot{\phi}+ \mathcalboondox{V}_{,\phi}=0. \label{General CR - P(X) equation of motion 3}
\end{equation}
Here, $c_s$ represents the sound speed which can be expressed as
\begin{equation}
	c_{s}^{2} = \frac{P_{, X}}{P_{, X}+2 X P_{, X X}}, \label{General CR - sound speed for P(X)}
\end{equation}
and the non-trivial field dynamics is controlled by the friction term, $3 c_{s}^{2} H \dot{\phi}$, and an effective potential given by
\begin{equation} 
	\mathcalboondox{V}_{,\phi}=\frac{c_s^2 \left(\dot{\phi}^2 P_{,X \phi}-P_{,\phi}\right)}{P_{,X}}.\label{effective potential-general expression} 
\end{equation}

In our first approach to add a constant-roll condition to the above system of equations, we consider the conventional form given by
\begin{equation}
	\frac{\ddot{\phi}}{H \dot{\phi}}= \beta_{\text{\tiny{F}}}, \label{General CR3 - constant-roll condition 3}
\end{equation}
where subscript F refers to a constant-roll condition that constrains the field rate of roll. We call this constraint the \textit{field-constant-roll} (FCR) condition. This constant-roll condition can be written as $\mathrm{~d} \ln \dot{\phi} / \mathrm{~d} N = \beta_{\text{\tiny{F}}}$, indicating that the scalar field has a constant roll rate. Integrating it out, we find
\begin{equation}
	\dot{\phi} \propto e^{\beta_{\text{\tiny{F}}} N}. \label{large gamma - general relation for dotphi}
\end{equation}
This relation indicates that $|\dot{\phi}|$ grows (decays) rapidly for $\beta_{\text{\tiny{F}}} > 0$ ($\beta_{\text{\tiny{F}}} < 0$). Note that this behavior is simply a consequence of the FCR condition and has been derived regardless of the form of the action.

As the second approach, we define a constant-roll condition that (similar to the canonical case) leads to a constant $\epsilon_{2}$. Provided that the first slow-roll parameter is small,\footnote{This is usual in the standard models of inflation based on inflationary attractor dynamics.} then the second slow-roll parameter $\epsilon_{2}  =2 \epsilon_{1}+ \ddot{H} / (H \dot{H})$ is roughly constant by imposing the following condition:
\begin{equation}
	\frac{\ddot{H}}{2 H \dot{H}}=\beta_{\text{\tiny{H}}}. \label{General CR1 - fist kind Contant roll condition}
\end{equation}
We refer to this constraint as the \textit{Hubble-constant-roll} (HCR) condition. Note that, unlike the canonical case, $\beta_{\text{\tiny{H}}}$ does not always equal to $\ddot{\phi}/(H \dot{\phi})$. The HCR condition does not ensure a constant rate of roll for the scalar field. Likewise, the FCR condition does not guarantee a constant $\epsilon_{2}$. An interesting property of the HCR condition is that we can directly integrate it to obtain the Hubble parameters given by \eqref{Canonical CR - hubble parameter 1}-\eqref{Canonical CR - hubble parameter 3}, and \eqref{Canonical CR - hubble parameter 0}. 
Accordingly, the slow-roll parameters and the acceptable ranges for $N$ and $\beta_{\text{\tiny{H}}}$ are also similar to the canonical case (unless the field becomes imaginary and imposes additional constraints).

To see how FCR and HCR conditions are interconnected and whether the latter can also be interpreted as a constant rate of roll, let us find the relation between what we call $\beta_{\text{\tiny{F}}}$ and $\beta_{\text{\tiny{H}}}$. We use equation \eqref{General CR - P(X) equation of motion 2} to write the FCR condition \eqref{General CR3 - constant-roll condition 3} as
\begin{equation}
	\frac{\ddot{H}}{H \dot{H}}=\frac{\dot{\phi} \, Q_{,X\phi}}{H Q_{,X}} + \left(\frac{c_s^2 + 1}{c_s^2} \right) \beta_{\text{\tiny{F}}}, \label{General CR3 - general condition between first and third CR}
\end{equation}
where we have defined the \textit{effective kinetic term}, $Q(X,\phi) \equiv P(X,\phi) + V(\phi)$. From this, we notice that when the FCR condition is imposed, $\ddot{H}/ (H \dot{H})$ is not necessarily constant. In a particular model, if the right-hand side of equation \eqref{General CR3 - general condition between first and third CR} becomes constant, the HCR condition is met. In this case, we can use \eqref{General CR1 - fist kind Contant roll condition} to rewrite \eqref{General CR3 - general condition between first and third CR} as 
\begin{equation}
	2 \beta_{\text{\tiny{H}}}=\frac{\dot{\phi} \,Q_{,X\phi}}{H Q_{,X}} + \left(\frac{c_s^2 + 1}{c_s^2} \right) \beta_{\text{\tiny{F}}}.\label{General CR3 - general relation between third and first CR}
\end{equation}
With fixed $P(X,\phi)$, for the models constrained with FCR and HCR conditions to be homologous, the condition \eqref{General CR3 - general relation between third and first CR} must be satisfied. Canonical constant-roll model satisfies this relation for $ \beta_{\text{\tiny{F}}}= \beta_{\text{\tiny{H}}}$.

A key property of the canonical constant-roll model is that it leads to USR for $\beta = -3$, and it can be considered a generalization of the USR model. However, our FCR and HCR conditions do not always correspond to the generalization of USR. In other words, in a particular model, these constant-roll conditions may not reduce to a flat USR potential for any choice of the constant-roll parameter. As our final approach to the generalized constant-roll  condition, we define the constant-roll conditions so that the profile of the scalar potential $V(\phi)$ is sufficiently flat when constant-roll parameter approaches $-3$. We approximate the field equation \eqref{General CR - P(X) equation of motion 3} in the USR limit as
\begin{equation}
	V_{,\phi} \ll \{\dot{\phi},\ddot{\phi}\}\quad \Rightarrow \quad \ddot{\phi}+3 c_{s}^{2} H \dot{\phi}+ \frac{c_s^2 \left(\dot{\phi}^2 Q_{,X \phi}-Q_{,\phi}\right)}{Q_{,X}}=0. \label{Three CR - USR limit for PX}
\end{equation}
According to this equation, we define our third constant-roll condition as
\begin{equation}
	\frac{\ddot{\phi}}{H\dot{\phi}}=c_s^2 \beta_{\text{\tiny{V}}}-\frac{c_s^2 \left(\dot{\phi}^2 Q_{,X\phi}-Q_{\phi}\right)}{H \dot{\phi} \, Q_{,X}}, \label{General CR3: constant-roll condition for P(X)}
\end{equation}
and refer to this constraint as the \textit{potential-constant-roll} (PCR) condition. Non-canonical models constrained by \eqref{General CR3: constant-roll condition for P(X)} are generalizations of USR, in the sense that they lead to a flat potential when $\beta_{\text{\tiny{V}}}$ approaches $-3$.
One may use equation \eqref{General CR - P(X) equation of motion 3} to rewrite the PCR condition \eqref{General CR3: constant-roll condition for P(X)} in the familiar form
\begin{equation}
	\dot{V}= 2 H \dot{H}(3+ \beta_{\text{\tiny{V}}}). \label{Three CR - PCR 2}
\end{equation}
Although our third manipulation of constant-roll constraint makes implicit use of the third expression in \eqref{canonic CR - constant-roll condition}, it
is important to emphasize that condition \eqref{Three CR - PCR 2} is not the only possible way to generalize the USR. For instance, the condition $\dot{V} = (3+ \beta_{\text{\tiny{V}}}) H V$, which again exhibits a flat potential when $\beta_{\text{\tiny{V}}} \rightarrow -3$,
can be viewed as another generalization of USR and may provide new insights into the model. Nevertheless, the form \eqref{Three CR - PCR 2} has two advantages: it is identical to FCR and HCR conditions in the canonical limit, 
and for the examples we will discuss later, $\beta_{\text{\tiny{V}}} \rightarrow 0$ leads to slow-roll case. Indeed, in this work, we discuss only a small number of an apparently endless variety of possible constant-roll conditions picked out by the demand of having a canonical counterpart.\footnote{The authors in \cite{motohashi2017f} have studied the constant-roll condition in $f(R)$ gravity by introducing the condition $\ddot {F}= \beta H_J \dot{ F}$ and $F \equiv df/dR$ to system of equations in Jordan frame. This condition, when transformed to Einstein frame through the relevant transformation relations, depends on the functional form of $f(R_J)$. Note that all three constant roll conditions in section \ref{sec: Three constant-roll Conditions} are independent of the functional form of $P(X,\phi)$; this will let us to talk about some general dynamical behaviors for generic $P(X,\phi)$ and compare the results. The generalization can be done in many ways and with different motivations \cite{motohashi2017f, motohashi2019constant, boisseau2019inflationary, oikonomou2022generalizing}. In reference \cite{motohashi2017f}, for example, simplicity and aesthetic elegance are stated as the reasons for the proposed condition. We concentrated on the conditions with canonical counterpart. In this work, we limit our analysis to $P(X,\phi)$ action, leaving the study of the modified gravity theories for future investigations.}


One can also derive the relation between $\beta_{\text{\tiny{V}}}$ and $\beta_{\text{\tiny{F}}}$ by utilizing definitions \eqref{General CR3 - constant-roll condition 3} and \eqref{General CR3: constant-roll condition for P(X)}
\begin{equation}
	\beta_{\text{\tiny{F}}}=c_s^2 \beta_{\text{\tiny{V}}}-\frac{c_s^2 \left(\dot{\phi}^2 Q_{,X\phi}-Q_{\phi}\right)}{H \dot{\phi} \, Q_{,X}}. \label{General CR3 - general relation between third and second CR}
\end{equation}
To find the relation between $\beta_{\text{\tiny{V}}}$ and $\beta_{\text{\tiny{H}}}$, we elaborate on equation \eqref{General CR - P(X) equation of motion 2} to rewrite the PCR condition \eqref{General CR3: constant-roll condition for P(X)} as
\begin{equation}
	\frac{\ddot{H}}{H \dot{H}}=\left(1+c_s^2\right) \beta_{\text{\tiny{V}}} +  \frac{ \left(c_s^2+1\right) Q_{,\phi}- c_s^2 \dot{\phi}^2 Q_{,X\phi}}{H \dot{\phi} \, Q_{,X}}. \label{General CR3: first CR from second CR P(X) case}
\end{equation}
This implies that given a non-canonical action, two constant-roll models based on PCR and HCR constraints would be homologous if we have
\begin{equation}
	2 \beta_{\text{\tiny{H}}}=\left(1+c_s^2\right) \beta_{\text{\tiny{V}}} +  \frac{ \left(c_s^2+1\right) Q_{,\phi}- c_s^2 \dot{\phi}^2 Q_{,X\phi}}{H \dot{\phi} \, Q_{,X}}.\label{General CR2 - relation between first and second CR}
\end{equation}
Finally, note that the system of equations \eqref{General CR3 - general relation between third and first CR}, \eqref{General CR3 - general relation between third and second CR}, and \eqref{General CR2 - relation between first and second CR} are dependent, in the sense that when every pair of the equations is satisfied, all three kinds of constant-roll conditions lead to similar results. The canonical constant-roll model satisfies these conditions for $\beta_{\text{\tiny{F}}} = \beta_{\text{\tiny{H}}} = \beta_{\text{\tiny{V}}}$. In what comes next, we arrange to clearly distinguish between the above approaches for generalizing the constant-roll conditions.

\section{Constant-Roll and Propagation Inflation} \label{sec: constant-roll and Propagation Inflation}
This section considers a special class of inflationary models with constant sound speed to demonstrate the applications of the constant-roll conditions discussed in section \ref{sec: Three constant-roll Conditions}. With constant $c_s$, the integration of equation \eqref{General CR - sound speed for P(X)} gives us
\begin{equation}
	P(X, \phi)=\mathcal{F}(\phi) X^{\lambda}-V(\phi), \label{General CR - Lagrangian for Cs=constant}
\end{equation}
where $\mathcal{F}(\phi)$ and $V(\phi)$ are two arbitrary functions of $\phi$, and for simplicity we have defined $\lambda \equiv X P_{,XX} / P_{,X}+1 = (c_s^2 + 1)/ (2 c_s^2) \geq
1$. We refer to this class of models as \textit{constant propagation inflation}. By imposing the constant-roll condition on the model in the following subsections, we will have a class of \textit{constant-roll and propagation inflation}. In the case of $c_s=1$, the constant propagation model \eqref{General CR - Lagrangian for Cs=constant} takes the form $P(X, \phi)= \mathcal{F} X - V$, also known as k/G inflation \cite{lin2020primordial}. 

For  the class of constant-roll and propagation models, equations \eqref{General CR - P(X) equation of motion 1}-\eqref{General CR - P(X) equation of motion 3} turn into
\begin{align}
	& 3 H^{2}=\frac{\mathcal{F}}{c_{s}^{2}}  X^{\lambda}+V, \label{General CR - equation of motion for Cs constant 1}\\
	& \dot{H}+ \lambda \mathcal{F}  X^{\lambda}=0, \label{General CR - equation of motion for Cs constant 2} \\
	&\ddot{\phi}+3 c_{s}^{2} H \dot{\phi}+\mathcalboondox{V}_{,\phi}=0. \label{General CR - equation of motion for Cs constant 3}
\end{align}
Here, the derivative of effective potential is given by
\begin{equation}
	\mathcalboondox{V}_{,\phi} = \frac{X \mathcal{F}_{, \phi}}{\lambda \mathcal{F}}+\frac{c_{s}^{2} V_{, \phi}}{\lambda \mathcal{F}}X^{1-\lambda}.
\end{equation}
For $c_s=\mathcal{F}=1$, we can verify that the general argument presented here leads to the canonical case discussed in section \ref{sec: Canonical CR Model}. 

In this section, we study three classes of constant-roll and propagation models constrained with constant-roll conditions introduced in the previous section. However, before proceeding, it is useful to determine whether these classes would be homologous or not. 
The effective kinetic term in the constant propagation model is given by
\begin{equation}
	Q= P + V=\mathcal{F} X^{\lambda}. \label{constant propagation - Q function}
\end{equation}
Using this, we can show that constant propagation models satisfy equation \eqref{General CR2 - relation between first and second CR} with $2 \beta_{\text{\tiny{H}}}=\left(1+c_s^2\right) \beta_{\text{\tiny{V}}}$. In other words, in a constant propagation model if $\beta_{\text{\tiny{H}}}$ is constant, $\beta_{\text{\tiny{V}}}$ will also be constant and vice versa. Therefore, HCR and PCR constrained classes are not independent and do not lead to different dynamics.
By knowing this, we can easily discuss the model's USR limit. For the HCR condition the Hubble parameter can be derived identical to the canonical case, and since in the constant propagation model the HCR and PCR condition are homologous, the Hubble parameter and slow-roll parameters in the USR case of the constant propagation models are given by equations \eqref{Canonical CR - hubble parameter 1} and \eqref{Canonical CR - SR parameters for solution 1}, but with $\beta $ replaced with $-3\left(c_{s}^{2}+1\right)/2$. As we discussed in section \ref{sec: Canonical CR Model}, if $\epsilon_{1}$ decays faster than $e^{-3N}$ modes keep growing at superhorizon scales. Thus, the model cannot be stable unless the condition $\beta = -3\left(c_{s}^{2}+1\right)/2 > -3/2$, or equivalently $c_s^2 <0 $, is satisfied, which is impossible. It follows that the USR limit of constant propagation model \eqref{General CR - Lagrangian for Cs=constant} can never be stable.

Using \eqref{constant propagation - Q function} again, we can write equation \eqref{General CR3 - general relation between third and first CR} as
\begin{equation}
	\beta_{\text{\tiny{H}}}=\frac{\dot{\mathcal{F}}}{2 H \mathcal{F}} + \lambda \beta_{\text{\tiny{F}}}. \label{constant propagation - relation between HCR and FCR}
\end{equation}
This means that FCR and HCR constrained classes are generally independent and lead to different solutions unless the offending non-universal term $\dot{\mathcal{F}} / (2 H \mathcal{F})$ is a constant. Therefore, for a general $\mathcal{F}$ function, we only need to consider one of the HCR and PCR conditions, and we will study the FCR condition separately.
\subsection{Field-Constant-Roll Condition} \label{sec: Field-Constant-Roll Condition}
We now discuss the dynamics of the constant propagation class of models \eqref{General CR - Lagrangian for Cs=constant} constrained with the FCR condition \eqref{General CR3 - constant-roll condition 3}. In this case, if $\mathcal{F}(\phi)$ is known, one can start the study by integrating the constant-roll condition \eqref{General CR3 - constant-roll condition 3} written in the following form:
\begin{equation}
	\beta_{\text{\tiny{F}}}=\frac{c_s^2}{H} \left[-\frac{2^{\lambda}  H_{, \phi}}{\lambda \mathcal{F}}\right]^{c_{s}^{2}} \left(\ln\frac{H_{, \phi}}{ \mathcal{F}}\right)_{, \phi}, \label{CP FCR - constant roll condition in terms of phi}
\end{equation}
where equation \eqref{General CR - equation of motion for Cs constant 2} is being put to use in the above relation. Having $\mathcal{F} (\phi)$, one can solve this equation to find $H(\phi)$. Once the Hubble parameter has been determined, the field velocity can be derived using equation \eqref{General CR - equation of motion for Cs constant 2} as follows:
\begin{equation}
	\dot{\phi} = \left[- \frac{2^\lambda H_{,\phi}}{\lambda \mathcal{F}} \right]^{c_s^2}, \label{FCR condition - general dotphi}
\end{equation}
and by utilizing equation \eqref{General CR - equation of motion for Cs constant 1} for the potential we easily find
\begin{equation}
	V= 3H^2 + \frac{H_{,\phi}}{\lambda c_s^2} \left[- \frac{2^\lambda H_{,\phi}}{\lambda \mathcal{F}} \right]^{c_s^2}. \label{FCR condition - analytical potential}
\end{equation}

With the help of equations \eqref{General CR3 - constant-roll condition 3} and \eqref{General CR - equation of motion for Cs constant 2}, for the first and second derivatives of effective potential we obtain
\begin{align}
	&\mathcalboondox{V}_{,\phi} = - (3 c_s^2 + \beta_{\text{\tiny{F}}}) H \dot{\phi} \label{CP FCR - dotVeff},\\
	&\mathcalboondox{V}_{,\phi \phi} = (3 c_s^2 + \beta_{\text{\tiny{F}}})  (\epsilon_{1} - \beta_{\text{\tiny{F}}})H^2.\label{CP FCR - ddotVeff}
\end{align}
Note that, contrary to the potential expression \eqref{FCR condition - analytical potential}, these relations are not directly dependent on $\mathcal{F}$ function. It is clear from equation \eqref{CP FCR - dotVeff} that the derivative of effective potential vanishes at zeros of $H$ and $\dot{\phi}$. On the other hand zeros of $H_{,\phi}$ and $\dot{\phi}$ coincide (see \eqref{FCR condition - general dotphi}) and we find the effective potential extrema at zeros of $H$ and ${H}_{,\phi}$. Based on equation \eqref{CP FCR - ddotVeff}, we notice that, like the canonical case, at extrema the concavity of effective potential is determined by $\beta_{\text{\tiny{F}}}$ values; it is concave for $\beta_{\text{\tiny{F}}} < -3 c_s^2$ and $\beta_{\text{\tiny{F}}} >0$, and is convex for $-3 c_s^2<\beta_{\text{\tiny{F}}} <0$. For $\beta_{\text{\tiny{F}}} = -3 c_s^2$ the effective potential is flat.

Using general relation \eqref{large gamma - general relation for dotphi} and equation \eqref{General CR - equation of motion for Cs constant 2}, for the number of $e$-folds we find
\begin{equation}
	N -N_{0} = \frac{c_s^2}{\beta_{\text{\tiny{F}}}} \ln \left(\frac{H_{,\phi}}{\mathcal{F}}\right). \label{CP FCR - general N}
\end{equation}
Let us assume that $H_{,\phi}$ has a zero at $\phi = \tilde{\phi}$. From equation \eqref{CP FCR - general N} we notice that as $\phi \rightarrow \tilde{\phi}$, the expression $\ln \left(H_{,\phi} / \mathcal{F} \right)$ diverges and $\beta_{\text{\tiny{F}}}\left(N -N_{0}\right)$ goes to negative infinity, if $\mathcal{F}(\tilde{\phi})\neq 0$. 
Using equations \eqref{General CR - equation of motion for Cs constant 2} and \eqref{CP FCR - general N} we can find the following general relations for the first and second slow-roll parameters:
\begin{align}
	&\epsilon_{1} \propto \frac{\mathcal{F}}{H^2} \, e^{2 \lambda \beta_{\text{\tiny{F}}} N}, \qquad\epsilon_{2}=2\epsilon_{1}+2\lambda \beta_{\text{\tiny{F}}} +\frac{\dot{\mathcal{F}}}{H\mathcal{F}}. \label{CP FCR - general e1 behavior}
\end{align}
As the scalar field moves toward the extrema of the effective potential at $\phi = \tilde{\phi}$, $\mathcal{F}(\phi)$ and $H(\phi)$ approach the constant values of $\mathcal{F}(\tilde{\phi})$ and $H(\tilde{\phi})$ and we have an exponentially decreasing (increasing)  $|\epsilon_{1}|$ for positive (negative) $\beta_{\text{\tiny{F}}}$. Furthermore, if the time scale of $\mathcal{F}$ variation is large enough (so that we can neglect the last term in $\epsilon_{2}$ expression) and $\epsilon_{1}$ is sufficiently small, for the second slow-roll parameter we get $\epsilon_{2} \approx 2 \lambda \beta_{\text{\tiny{F}}} $. Therefore, the model exhibits a \textit{quasi-canonical} behavior in the vicinity of $\tilde{\phi}$ (in the sense that at this limit slow-roll parameters behave like the canonical case while the action is not canonical. If we have $\mathcal{F}(\tilde{\phi}) = 1$, then we will observe normal canonical behavior). If this
phase is characterized by a large negative $\tilde{\beta}_{\text{\tiny{F}}} \equiv \lambda \beta_{\text{\tiny{F}}}  < -3/2$, its dynamics
enter the non-attractor era lasting some $e$-folds of evolution during which the first slow-roll parameter decays exponentially.

The solutions with $\beta_{\text{\tiny{F}}} < -3 c_s^2$ are of phenomenological interest. In this case, the scalar field rolls up a bump-like region and asymptotically comes to rest at the top of the bump. This will be the case for the fine-tuned initial condition and these constant-roll solutions lead classically to eternal inflation. Under quantum perturbations, however, this will be a transient solution.
As we will show in an example in section \ref{sec: example for CP model}, the generic attractor solution is the dual case $\beta_{\text{\tiny{F}}}\rightarrow -(\beta_{\text{\tiny{F}}}+3c_s^2)$ (which for $c_s=1$, it gives rise to the duality relation discussed in \cite{morse2018large, lin2019dynamical} for canonical case).
The transition from the smaller $\beta_{\text{\tiny{F}}}$ to the larger value $-(\beta_{\text{\tiny{F}}}+3c_s^2)$ can be seen through the Klein-Gordon equation \eqref{General CR - equation of motion for Cs constant 3} linearized around the maximum of the effective potential,
\begin{equation}
	\ddot{\phi}+3 c_s^2 H(\tilde{\phi}) \dot{\phi}-\beta_{\text{\tiny{F}}}(\beta_{\text{\tiny{F}}}+3c_s^2)H^2 (\tilde{\phi}) (\phi - \tilde{\phi})\approx 0, \label{FCR expansion of KG equation}
\end{equation}
which is invariant under the duality transform. By utilizing \eqref{CP FCR - dotVeff}, an equivalent definition of  $\beta_{\text{\tiny{F}}}$ is given by $\beta_{\text{\tiny{F}}}=-3 c_s^2-\mathcalboondox{V}_{,\phi} / (H\dot{\phi})$. This form indicates that (if the field velocity is large enough so that the field can pass the extrema) when the field evolves through the critical point $\phi=\tilde{\phi}$, the parameter $\beta_{\text{\tiny{F}}}$ is forced to $-3c_s^2$.  The value of $-3c_s^2$ for the $\beta_{\text{\tiny{F}}}$ at the maximum of the effective potential is a feature of FCR condition,
and as we will show later, it will depend on the $\mathcal{F}$ profile in the HCR (and equivalently PCR) models.

\subsection{Hubble-Constant-Roll and Potential-Constant-Roll Conditions} \label{sec: second-kind constant-roll condition}
We discussed that in the constant propagation class \eqref{General CR - Lagrangian for Cs=constant} the HCR and PCR conditions are analogous. Therefore, we only consider HCR condition \eqref{General CR1 - fist kind Contant roll condition} here. The Hubble parameter in these models is similar to the canonical case. Accordingly, the first and second slow-roll parameters are given by 
\begin{equation}
	\epsilon_{1} \propto \frac{1}{H^2} \, e^{2 \beta_{\text{\tiny{H}}} N}, \qquad \epsilon_{2}=2\epsilon_{1}+2 \beta_{\text{\tiny{H}}}. \label{CP PCR - general e1 behavior}
\end{equation}
We observe that in general slow-roll parameters here differ from \eqref{CP FCR - general e1 behavior} parameters derived using the FCR condition. However, at the critical point $\tilde{\phi}$, both HCR and FCR constrained models exhibit an approximately identical quasi-canonical behavior with $\beta_{\text{\tiny{H}}} =\tilde{\beta}_{\text{\tiny{F}}} \equiv \lambda \beta_{\text{\tiny{F}}}$. Indeed, this is well consistent with our previous general argument that in constant propagation models three constant-roll conditions are homologous when the
timescale of $\mathcal{F}$ variation is large (see equation \eqref{constant propagation - relation between HCR and FCR}).


After removing $\mathcal{F}$ between equations \eqref{General CR - equation of motion for Cs constant 1} and \eqref{General CR - equation of motion for Cs constant 2} for the potential will have
\begin{equation}
	V=3 H^{2}+\frac{2 \dot{H}}{c_{s}^{2}+1}. \label{General CR1 - equation for V}
\end{equation}
In this equation, we can write the time derivative in terms of $N$ and use Hubble parameters \eqref{Canonical CR - hubble parameter 1}-\eqref{Canonical CR - hubble parameter 3}, and \eqref{Canonical CR - hubble parameter 0} to find the following potentials respectively:
\begin{align}
	&V(N)= M^2 \left[3+\left(\frac{2 \beta_{\text{\tiny{H}}}}{c_s^2+1}+3\right) e^{2 \beta_{\text{\tiny{H}}} N }\right], \label{General CR1 - potential 1}\\
	&V(N)= M^2 \left[3-\left(\frac{2 \beta_{\text{\tiny{H}}}}{c_s^2+1}+3\right) e^{2 \beta_{\text{\tiny{H}}} N }\right], \qquad \beta_{\text{\tiny{H}}} > 0, \label{General CR1 - potential 2} \\
	&V(N)= M^2 \left[\left(\frac{2 \beta_{\text{\tiny{H}}}}{c_s^2+1}+3\right) e^{2 \beta_{\text{\tiny{H}}} N }-3\right], \qquad \beta_{\text{\tiny{H}}} < 0,\label{General CR1 - potential 3}\\
	&V(N)= M^2 \left[3 N + \frac{1}{c_s^2+1}\right]. \label{General CR1 - potential 4}
\end{align}
The first three potentials are flat for $\beta_{\text{\tiny{H}}}=0$ and $\beta_{\text{\tiny{H}}} = -3\left(c_{s}^{2}+1\right)/2$. When $\beta_{\text{\tiny{H}}}=0$, all slow-roll parameters are small, so this parameter value corresponds to a slow-roll case.
By the homologous relation mentioned above and described by $2\beta_{\text{\tiny{H}}} = (c_s^2+1)\beta_{\text{\tiny{V}}}$, the latter appears to be an USR case.

With Hubble parameter, potential, and slow-roll parameters in hand, we need to know the form of the parameter $\mathcal{F}$ to find the dynamics of the inflaton field $\phi(t)$ (or $\phi(N)$). One can use equation \eqref{General CR - equation of motion for Cs constant 2} and its derivative to rewrite the constant-roll condition \eqref{General CR1 - fist kind Contant roll condition} as
\begin{equation}
	\beta_{\text{\tiny{H}}}=  \left[\frac{- 2^\lambda H_{, \phi}}{\lambda \mathcal{F}}\right]^{c_{s}^{2}} \left[\frac{\lambda c_s^2}{H} \left(\ln \frac{H_{,\phi}}{F}\right)_{,\phi} + \frac{\mathcal{F}_{,\phi}}{2 H \mathcal{F}}\right], \label{CP HCR - constant roll condition in terms of phi}
\end{equation}
and integrate it out to find $H(\phi)$. The field is then calculated using the relation $H(\phi)=H(t)=H(N)$.

The effective potential gradient can be obtained from the general expression \eqref{effective potential-general expression} and the HCR condition \eqref{General CR1 - fist kind Contant roll condition} in the form of
\begin{equation}
	\mathcalboondox{V}_{,\phi}= c_s^2 H\dot{\phi}\left[\frac{\dot{\mathcal{F}}}{2 \lambda c_s^2 H\mathcal{F}}-\left(3 + \beta_{\text{\tiny{V}}} \right)\right].\label{effective potential-PCR}
\end{equation}
The effective potential has extrema at $\phi=\tilde{\phi}$, i.e, where $\dot{\phi}$ or $H$ vanishes. The expression for the second derivative of the effective potential is quite complex but at $\phi = \tilde{\phi}$ it can be approximated as
\begin{equation}
	\mathcalboondox{V}_{,\phi \phi} \approx - c_s^4 H^2(\tilde{\phi}) \beta_{\text{\tiny{V}}} \left(3 + \beta_{\text{\tiny{V}}}\right). \label{second derivarive of effective potential-PCR}
\end{equation}
This term appears in the linearized Klein-Gordon equation around $\phi = \tilde{\phi}$. It is clear that at the critical point $\tilde{\phi}$ expressions \eqref{effective potential-PCR} and \eqref{second derivarive of effective potential-PCR} become similar to those in the FCR constrained models,  \eqref{CP FCR - dotVeff} and \eqref{CP FCR - ddotVeff}, with $\beta_{\text{\tiny{H}}} = \lambda c_s^2 \beta_{\text{\tiny{V}}}= \lambda \beta_{\text{\tiny{F}}}$. Therefore, based on our analysis, we expect to encounter the duality between $\beta_{\text{\tiny{H}}}$ and $-(\beta_{\text{\tiny{H}}}+ 3\lambda c_{s}^{2}) $ solutions. Nevertheless, this discussion does not show that the model will actually transition to the dual solution. To determine whether $\beta_{\text{\tiny{H}}}$ is parametrically forced to $-(\beta_{\text{\tiny{H}}}+ 3\lambda c_{s}^{2})$, we need to conduct a stability analysis that considers perturbations in the initial values of the model. We will perform such an analysis in the example provided in section \ref{sec: example for CP model}.  Indeed, approximation \eqref{FCR expansion of KG equation} only holds at the beginning and end of the transition trajectory, where the model approximately follows a constant-roll scenario. Hence, the transition trajectory cannot be accurately described by this approximation. The form of the transition trajectory generally depends on the specific form of the action.  As a simple demonstration of this dependence, we can use equation \eqref{effective potential-PCR} to express the constant-roll parameter as $\beta_{\text{\tiny{V}}} = \dot{\mathcal{F}} / (2 \lambda c_s^2 H\mathcal{F}) - \mathcalboondox{V}_{,\phi}/(c_s^2 H\dot{\phi}) + 3 $.  This relation highlights that at the extrema of the effective potential, where $\mathcalboondox{V}_{,\phi} = 0$, parameter $\beta_{\text{\tiny{V}}}$ depends on the specific form of the parameter $\mathcal{F}$ (unlike in the FCR case, where it is always equal to $-3c_s^2$ irrespective of the form of the $\mathcal{F}$ function and field velocity).

Using relations \eqref{General CR1 - fist kind Contant roll condition} and \eqref{General CR - equation of motion for Cs constant 2}, number of $e$-folds and the field are related by
\begin{equation}
	N-N_0=\frac{1}{2\beta_{\text{\tiny{H}}}}\left[ 2 \lambda c_s^2 \ln\left(\frac{H_{,\phi}}{\mathcal{F}} \right) + \ln \mathcal{F} \right].\label{number of efolds-HCR}
\end{equation}
It is also worth noting that with constant $c_{s}$ and slow-roll parameters similar to canonical models, the spectral index is also the same as the canonic (see Appendix \ref{Appendix: Power Specturm}). 

\subsection{An Example: k/G Inflation} \label{sec: example for CP model}

Having constructed the framework of constant-roll generalizations in constant propagation inflation, we will examine a particular example to demonstrate the mechanism discussed in sections \ref{sec: Field-Constant-Roll Condition} and \ref{sec: second-kind constant-roll condition}. We consider a subset of \eqref{General CR - Lagrangian for Cs=constant} as
\begin{equation}
	P(X, \phi)=\mathcal{F}(\phi) X -V(\phi) , \label{Example - action}
\end{equation}
where we have chosen $c_s = \lambda = 1$. These models are well motivated and studied in \cite{lin2020primordial}. For this reason, we found it interesting to study its application in constant-roll inflationary models. 
For $c_s=1$ general relations \eqref{CP FCR - constant roll condition in terms of phi} and \eqref{CP HCR - constant roll condition in terms of phi} become
\begin{align}
	&\left(\frac{H_{,\phi}}{\mathcal{F}}\right)_{,\phi}+\frac{\beta_{\text{\tiny{F}}}}{2}H=0,\label{Example CP - FCR condition} \\
	&\left(\frac{H_{,\phi}}{\sqrt{\mathcal{F}}}\right)_{,\phi}+\beta_{\text{\tiny{H}}}\sqrt{\mathcal{F}}H=0.\label{Example CP - HCR condition}
\end{align}
We recall that these equations are the FCR and HCR conditions written in terms of derivatives with respect to the field. These equations are in the form of Sturm–Liouville equations with eigenvalues given by constant-roll parameters and orthogonal eigenfunctions with zeros at $\phi=\tilde{\phi}$. Note that in equation \eqref{Example CP - HCR condition}, $\beta_{\text{\tiny{H}}}$ can be replaced by $\beta_{\text{\tiny{V}}}$. As we can see, the Hubble parameter $H(\phi)$ is determined by the function $\mathcal{F(\phi)}$. For definiteness, we focus on a representative example, $\mathcal{F}=\gamma \phi$, and discuss its properties. Here, $\gamma$ is a parameter with unit length dimension. With this choice, the following solutions are obtained for equations \eqref{Example CP - FCR condition} and \eqref{Example CP - HCR condition}, respectively:
\begin{align}
	H(\phi) =& C_{1} A i^{\prime} \left( \sqrt[3]{-\frac{\gamma \beta_{\text{\tiny{F}}}}{2}} \phi \right) + C_2 B i^{\prime} \left( \sqrt[3]{-\frac{\gamma \beta_{\text{\tiny{F}}}}{2}} \phi \right), \label{Example CP - solution for FCR conditions}\\
	H(\phi) =& C_3 \cosh \left(\frac{1}{3} \sqrt{-2 \gamma \beta_{\text{\tiny{H}}} \phi^3} \right) + C_4 \sinh \left(\frac{1}{3} \sqrt{-2 \gamma \beta_{\text{\tiny{H}}} \phi^3} \right), \label{Example CP - solution for HCR conditions}
\end{align}
where $C_i$s are integration constants, $A i (z)$ and $B i (z)$ are Airy functions of the first and second kind, and prime denotes the derivative with respect to $\phi$.

\begin{figure}
	\centering
	\begin{tikzpicture}
		\node[align=left] (img) at (0,5.7)
		{\quad\includegraphics{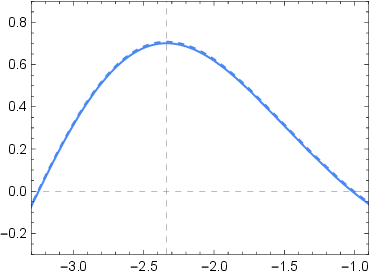}\quad\quad\quad\quad\includegraphics{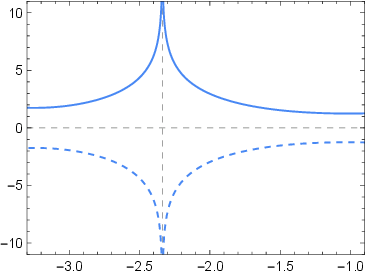}\quad};
		\node[align=left] (img) at (0,0)
		{\quad\,\,\includegraphics{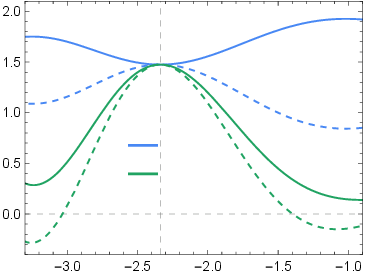}\quad\quad\quad\quad\,\includegraphics{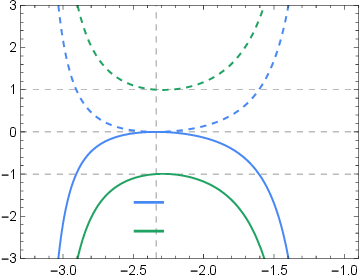}\quad}; //
		
		\node at (-7.2,5.88) {{\small \rotatebox{90}{$H/M$}}};
		\node at (+0.65,5.88) {{\small \rotatebox{90}{$N-N_0$}}};
		
		\node at (-3.65,-2.6) {{\small $\phi_{\text{\tiny{F}}}$}};
		\node at (+4.05,-2.6) {{\small $\phi_{\text{\tiny{F}}}$}};
		\node at (-3.65,+3.1) {{\small $\phi_{\text{\tiny{F}}}$}};
		\node at (+4.05,+3.1) {{\small $\phi_{\text{\tiny{F}}}$}};
		
		\node at (-3.74,-0.18) {{\scriptsize $\mathcalboondox{V}/M^2$}};
		\node at (-3.74,-0.68) {{\scriptsize $V/M^2$}};
		
		\node at (+3.84,-1.1) {{\scriptsize $\epsilon_{1}$}};
		\node at (+3.84,-1.58) {{\scriptsize $\epsilon_{2}$}};
	\end{tikzpicture}
	\caption{The plots of the Hubble parameter \eqref{Example CP - hubble case 1} (top left panel), number of $e$-folds \eqref{CP Example - Number e-fold} (top right panel), potential \eqref{CP Example - potential} and effective potential \eqref{CP Example - effective potential} (bottom left panel), and slow-roll parameters \eqref{CP Example - slow-roll parameter 1} and \eqref{CP Example - slow-roll parameter 2} (bottom right panel), once for $\beta_{\text{\tiny{F}}} = -0.5$ and $\gamma=1$ (solid lines) and once for $\beta_{\text{\tiny{F}}} = 0.5$ and $\gamma=-1$ (dashed lines). The parameter $\mathcalboondox{V}_0$ in effective potential is taken to be $1.475$. Vertical grid lines in all panels correspond to the extrema of $H_{,\phi}$ at $\phi_{\text{\tiny{F}}} =  a_1 \approx -2.34$.}
	\label{fig: F = gamma x phi}
\end{figure}

Let us first examine the FCR solution \eqref{Example CP - solution for FCR conditions}. For convenience, we consider a particular case with $C_2 = 0$ as
\begin{equation}
	H(\phi_{\text{\tiny{F}}}) = M A i^{\prime} (\phi_{\text{\tiny{F}}}) . \label{Example CP - hubble case 1}
\end{equation}
Here, we have defined $M \equiv C_1$, $ \bar{\beta}_{\text{\tiny{F}}}\equiv  \sqrt[3]{-\beta_{\text{\tiny{F}}} \gamma / 2}$, and $\phi_{\text{\tiny{F}}} \equiv \bar{\beta}_{\text{\tiny{F}}} \phi$.
A plot of the Hubble parameter \eqref{Example CP - hubble case 1}  is presented in Figure \ref{fig: F = gamma x phi} (top left panel). To have a real Hubble parameter, we need to choose $\gamma \beta_{\text{\tiny{F}}} < 0$. As we wish to examine the model for both positive and negative $\beta_{\text{\tiny{F}}}$, we choose $\beta_{\text{\tiny{F}}} = -0.5$ with $\gamma = 1$, as well as  $\beta_{\text{\tiny{F}}} = 0.5$ with $\gamma = -1$. These two cases share the same Hubble parameter but have different dynamics. Note that with these choices, $\bar{\beta}_{\text{\tiny{F}}}$ is positive and zeros of $H_{,\phi} \propto \phi_{\text{\tiny{F}}} A i (\phi_{\text{\tiny{F}}})$ are located at $\phi_{\text{\tiny{F}}} = 0$ and isolated values of $\phi_{\text{\tiny{F}}}$ on negative real axis.
Following the notation of reference \cite{olivier2010airy}, we show $s^{\text{th}}$ zeros of $Ai(z)$ and $Ai^{\prime} (z)$ with $a_s$ and $a^\prime_s$, respectively.
Specifically, we will focus on the model around the $\phi_{\text{\tiny{F}}} = a_1 \approx -2.34$ point where $H_{,\phi} = 0$ and quasi-canonical behavior is expected. To avoid any ambiguity, we further restrict our discussion to the $a^{\prime}_1 < \phi_{\text{\tiny{F}}} < a^{\prime}_2$ range, where the Hubble parameter remains positive for $M > 0$. As we will demonstrate later, the inflaton naturally resides within this region, as inflation ends before the sign of Hubble parameter changes.

Having the Hubble parameter, we can use the general relation \eqref{CP FCR - general N} to find the number $e$-folds as
\begin{equation}
	N (\phi_{\text{\tiny{F}}}) = \frac{1}{\beta_{\text{\tiny{F}}}} \ln \left( \left|  Ai \left( \phi_{\text{\tiny{F}}}  \right) \right|\right) + N_0. \label{CP Example - Number e-fold}
\end{equation}
The plot of $N(\phi_{\text{\tiny{F}}})$ can be seen in the top right panel of Figure \ref{fig: F = gamma x phi}.
We see that $N(\phi_{\text{\tiny{F}}})$ is singular at $\phi_{\text{\tiny{F}}} = a_1$, where  $H_{,\phi} \rightarrow 0$, as expected.

To derive the potential we can use equations \eqref{FCR condition - analytical potential} to get
\begin{equation}
	V= M^2 \left[ \beta_{\text{\tiny{F}}}    \phi_{\text{\tiny{F}}} Ai(\phi_{\text{\tiny{F}}} )^2+3 Ai'(\phi_{\text{\tiny{F}}} )^2 \right]. \label{CP Example - potential}
\end{equation}
Equations \eqref{FCR condition - general dotphi} and \eqref{CP FCR - dotVeff} give the gradient of effective potential in the form of
\begin{equation}
	\mathcalboondox{V}_{,\phi}=  - \frac{M^2 \beta_{\text{\tiny{F}}}  (\beta_{\text{\tiny{F}}} +3)}{\bar{\beta}_{\text{\tiny{F}}}}   Ai(\phi_{\text{\tiny{F}}}) Ai'(\phi_{\text{\tiny{F}}}).
\end{equation}
Fortunately, we can also integrate this relation to find the effective potential as 
\begin{equation}
	\mathcalboondox{V} = \frac{M^2 \bar{\beta}_{\text{\tiny{F}}} (\beta_{\text{\tiny{F}}} +3) }{\gamma }  Ai(\phi_{\text{\tiny{F}}})^2 + \mathcalboondox{V}_0, \label{CP Example - effective potential}
\end{equation}
where $\mathcalboondox{V}_0$ is the constant of integration. Potential \eqref{CP Example - potential} and effective potential \eqref{CP Example - effective potential} are plotted in Figure \ref{fig: F = gamma x phi} (bottom left panel). The effective potential
shows a roller-coaster profile with plateau-like regions connected to steep cliffs. Effective potential extrema are located at $a_s$ and $a^\prime_s$ which correspond to zeros of $H_{,\phi}$ and $H$, respectively. We recall that in FCR models the effective potential is not directly related to $\mathcal{F}$ and its concavity at extrema depends only on the value of the constant-roll parameter $\beta_{\text{\tiny{F}}}$. However, this is not the case for the potential. In Figure \ref{fig: F = gamma x phi} we see that for $\beta_{\text{\tiny{F}}} = -0.5$ the effective potential is convex and for $\beta_{\text{\tiny{F}}} = 0.5$ it is concave, while the potential is concave in both cases.

Depending on the field's initial value,\footnote{The dynamics of the constant propagation models can be completely determined by the field's initial value as the field velocity is evaluated by equation \eqref{FCR condition - general dotphi} and we do not have freedom of choosing it.} one can think of different inflationary scenarios. In the case of $\beta_{\text{\tiny{F}}} = -0.5$, the singularity of $N(\phi_{\text{\tiny{F}}})$ plot at $a_1$ shows that if the analytic field (that is described by constant $\beta_{\text{\tiny{F}}}$) starts at the neighbourhood of $\phi_{\text{\tiny{F}}} = a_1$, it will eventually rest at extrema.
The field in $\beta_{\text{\tiny{F}}} = 0.5$ case always rolls in the opposite direction to the $\beta_{\text{\tiny{F}}} = -0.5$ case.

\begin{figure}
	\centering
	\begin{tikzpicture}
		\node[align=left] (img) at (0,0)
		{\quad\,\,\includegraphics{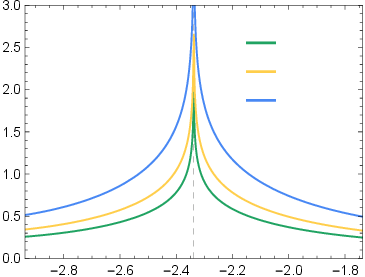}\quad\quad\quad\quad\,\includegraphics{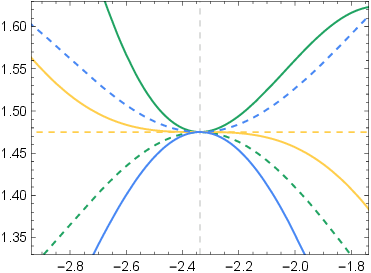}\quad}; //
		
		\node at (-7.25,0.2) {{\small \rotatebox{90}{$N-N_0$}}};
		
		\node at (-3.65,-2.6) {{\small $\phi_{\text{\tiny{F}}}$}};
		\node at (+4.27,-2.6) {{\small $\phi_{\text{\tiny{F}}}$}};
		
		\node at (-1.78,1.60) {{\tiny $\beta_{\text{\scalebox{.8}{F}}} = -4$}};
		\node at (-1.78,1.13) {{\tiny $\beta_{\text{\scalebox{.8}{F}}} = -3$}};
		\node at (-1.78,0.64) {{\tiny $\beta_{\text{\scalebox{.8}{F}}} = -2$}};
		
	\end{tikzpicture}
	\caption{Another plot of the number of $e$-folds \eqref{CP Example - Number e-fold} (left panel), potential \eqref{CP Example - potential} (solid lines in right panel), and effective potential \eqref{CP Example - effective potential} (dashed lines in right panel) for $\gamma = 1$ and $\mathcalboondox{V}_0 = 1.475$. Vertical grid line at $\phi_{\text{\tiny{F}}}=a_1$ corresponds to the singularity of function $N(\phi_{\text{\tiny{F}}})$.}
	\label{fig: F = gamma x phi 2}
\end{figure}

Next, we derive and analyse slow-roll parameters. Using equations \eqref{FCR condition - general dotphi} and \eqref{Example CP - hubble case 1}, for the first and second slow-roll parameters we find
\begin{align}
	\epsilon_{1} &= -\beta_{\text{\tiny{F}}} \phi_{\text{\tiny{F}}}   \left(\frac{Ai(\phi_{\text{\tiny{F}}})}{Ai'(\phi_{\text{\tiny{F}}})}\right)^2, \label{CP Example - slow-roll parameter 1}\\
	\epsilon_{2} &= \beta_{\text{\tiny{F}}} \left[2 + \frac{1}{\phi_{\text{\tiny{F}}}}  \left( \frac{ Ai(\phi_{\text{\tiny{F}}})}{ Ai'(\phi_{\text{\tiny{F}}})}\right) -2 \phi_{\text{\tiny{F}}}  \left(\frac{Ai(\phi_{\text{\tiny{F}}})}{Ai'(\phi_{\text{\tiny{F}}})}\right)^2 \right]. \label{CP Example - slow-roll parameter 2}
\end{align}
Figure \ref{fig: F = gamma x phi} (bottom right panel) illustrates these parameters.
At $\phi_{\text{\tiny{F}}} = a_1$, $\epsilon_{1}$ is approaching zero, while $\epsilon_{2}$ is roughly $2 \beta_{\text{\tiny{F}}}$. When $\beta_{\text{\tiny{F}}}=-0.5$, the field moves toward the extrema of effective potential, where we observe quasi-canonical behavior, whereas when $\beta_{\text{\tiny{F}}}=0.5$, the field moves away from this point, which means we gradually leave the quasi-canonical phase. When $\beta_{\text{\tiny{F}}}$ is positive, $\epsilon_{1}$ grows in time and we can find points where the first slow-roll parameter reaches 1 and inflation ends. Additionally, since we have $\epsilon_{2} \approx 2 \beta_{\text{\tiny{F}}}$ when $\phi_{\text{\tiny{F}}} \rightarrow a_1$, choosing $\beta_{\text{\tiny{F}}} < -1$ can potentially lead to a burst in the power spectrum and the formation of primordial black holes. We will talk more about this mechanism in another example in section \ref{sec: k inflation - PBH formation}.

In Figure \ref{fig: F = gamma x phi} we showed how the concavity of potential and effective potential changes around $\beta_{\text{\tiny{F}}} = 0$. In Figure \ref{fig: F = gamma x phi 2} (right panel) we plotted the potential and effective potential again, this time for $\beta_{\text{\tiny{F}}} = -2$, $-3$, and $-4$, around the critical point $\phi_{\text{\tiny{F}}} = a_1$, to show how the concavity changes around $\beta_{\text{\tiny{F}}} = -3$. We can see that the potential is convex for $\beta_{\text{\tiny{F}}} = -4$ and concave for $\beta_{\text{\tiny{F}}} = -2$. In each case, since $\mathcal{F}$ is negative, the concavity of effective potential is in the opposite direction of the potential. When $\beta_{\text{\tiny{F}}} = -3$, we have $\mathcalboondox{V}_{,\phi \phi} = \mathcalboondox{V}_{,\phi }=0$ and effective potential is flat throughout. There is, however, only one point (at the singularity of $N(\phi_{\text{\tiny{F}}})$) where the first and second derivatives of potential are zero. Actually, we do not have a flat USR-like potential for any choice of the constant-roll parameter $\beta_{\text{\tiny{F}}}$. To study the USR limit of this model, one should use the PCR condition or equivalently the HCR condition.

\begin{figure}
	\centering
	\begin{tikzpicture}
		\node[align=left] (img) at (0,0)
		{\quad\,\,\includegraphics{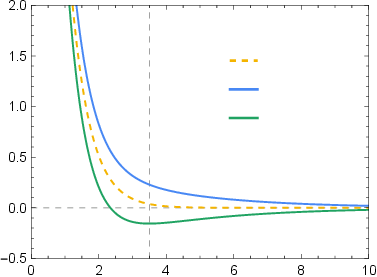}\quad\quad\quad\quad\,\includegraphics{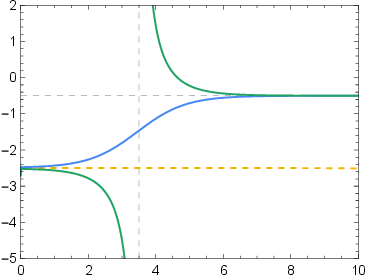}\quad}; //
		
		\node at (-7.25,0.2) {{\small \rotatebox{90}{$(\phi-a_1/\bar{\beta}_{\text{\tiny{F}}}) \times 10^2$}}};
		\node at (+0.57,0.2) {{\small \rotatebox{90}{$\beta_{\text{\tiny{F}}}$}}};
		
		\node at (-3.65,-2.6) {{\small $t$}};
		\node at (+4.27,-2.6) {{\small $t$}};
		
		\node at (-2.25,1.31) {{\scriptsize $\delta = 0$}};
		\node at (-1.95,0.84) {{\scriptsize $\delta = -0.01$}};
		\node at (-1.95,0.35) {{\scriptsize $\delta = +0.01$}};
		
	\end{tikzpicture}
	\caption{We have numerically integrated the field equation \eqref{General CR - equation of motion for Cs constant 3} for $\beta_{\text{\tiny{F}}}=-2.5$, $\gamma=1$, and $M=1$, to evaluate the time evolution of the field (left panel) and the constant-roll parameter $\beta_{\text{\tiny{F}}} = \ddot{\phi} / (H \dot{\phi})$ (right panel), in order to assess the stability of the constant-roll and propagation model \eqref{Example CP - hubble case 1} under small perturbations of field velocity, $\delta$. Initial values for the field and its velocity are taken as $-2$ and $-0.295-\delta$, respectively. Vertical grid lines correspond to the time when field stops and reverses its direction in the $\delta = +0.01$ case. The horizontal grid line in the right panel shows the corresponding dual attractor solution $\beta_{\text{\tiny{F}}} = -0.5$.}
	\label{fig: stabiliuy analysis}
\end{figure}
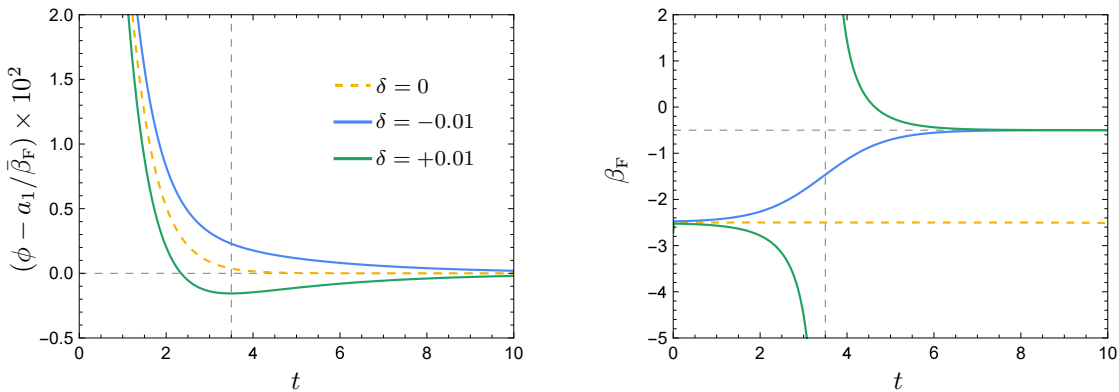

To assess the stability of the model, we employ the methodology outlined in reference \cite{lin2019dynamical}, imposing a small perturbation that deviates the field evolution from the analytic solution corresponding to a constant $\beta_{\text{\tiny{F}}}$. The evolution of the field and deviation of constant-roll parameter $\beta_{\text{\tiny{F}}}$ from the initially set value under small perturbations in the field velocity are plotted in Figure \ref{fig: stabiliuy analysis}. We observe that in the unperturbed constant-roll case, the field ultimately settles at the extrema of the effective potential, with the constant-roll parameter $\beta_{\text{\tiny{F}}} = -2.5$ remaining unchanged throughout. However, when the absolute value of the initial field velocity $|\dot{\phi}_0|$ is slightly reduced, the model promptly transitions to the corresponding dual attractor solution $\beta_{\text{\tiny{F}}}=-0.5$ (see \eqref{FCR expansion of KG equation}) and then settles at the extrema. The situation is a little different when we slightly increase $|\dot{\phi}_0|$. In this case, as the field reaches the extrema of the effective potential, where $\beta_{\text{\tiny{F}}}$ approaches $-3$, it possesses sufficient kinetic energy to surpass this point. After surpassing the extrema, the field velocity gradually diminishes until it completely stops and reverses its direction. The model then immediately falls into the attractor dual solution $\beta_{\text{\tiny{F}}}=-0.5$ and eventually comes to rest at the extrema.

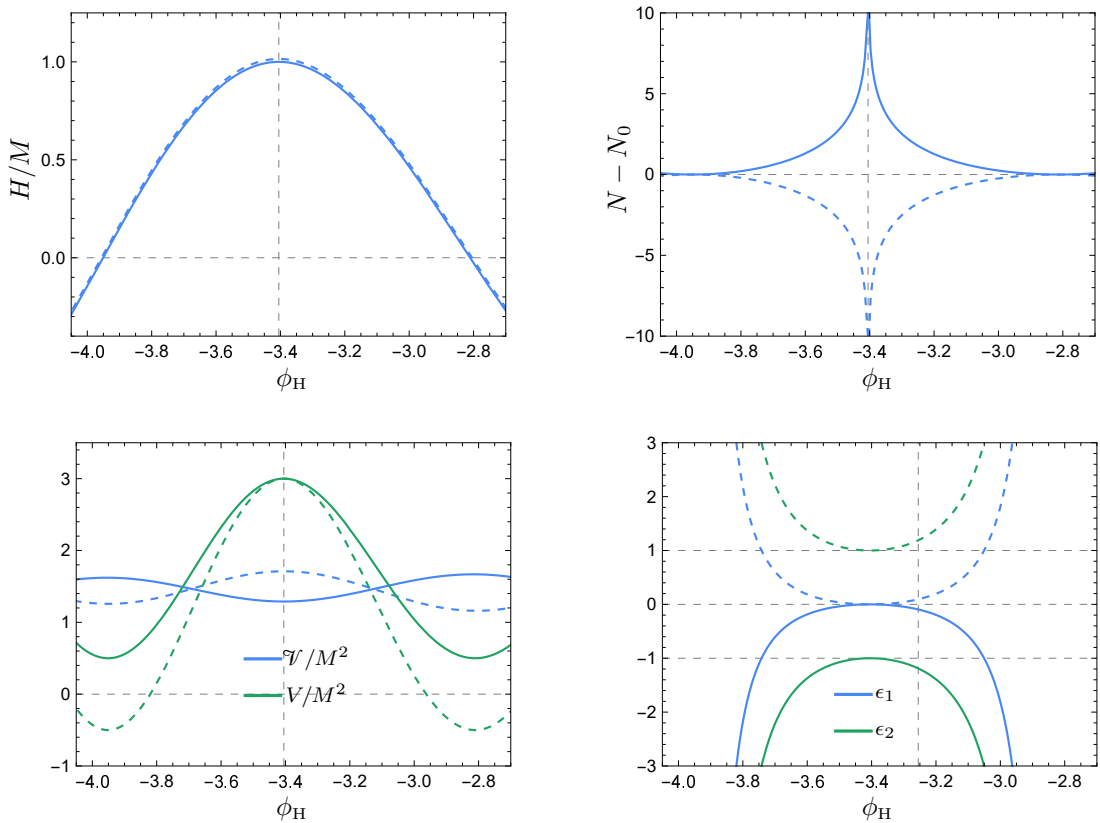
\begin{figure}
	\centering
	\begin{tikzpicture}
		\node[align=left] (img) at (0,5.7)
		{\quad\includegraphics{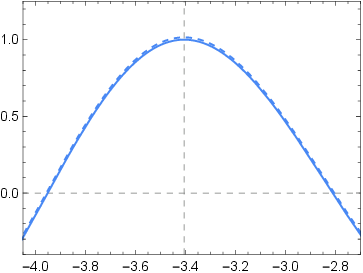}\quad\quad\quad\quad\includegraphics{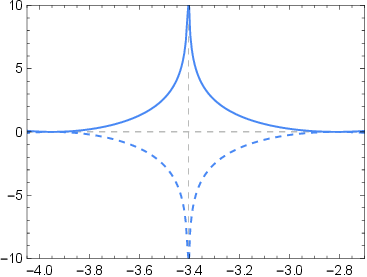}\quad};
		\node[align=left] (img) at (0,0)
		{\quad\,\,\includegraphics{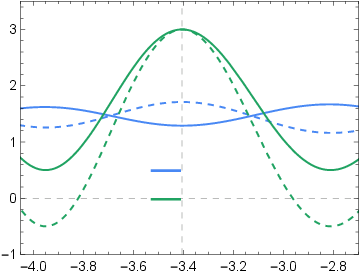}\quad\quad\quad\quad\,\includegraphics{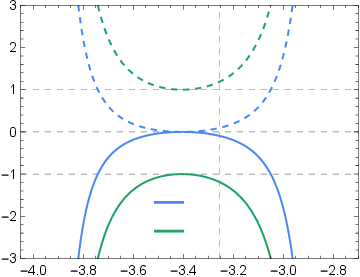}\quad}; //
		
		\node at (-7.2,5.88) {{\small \rotatebox{90}{$H/M$}}};
		\node at (+0.65,5.88) {{\small \rotatebox{90}{$N-N_0$}}};
		
		\node at (-3.65,-2.6) {{\small $\phi_{\text{\tiny{H}}}$}};
		\node at (+4.05,-2.6) {{\small $\phi_{\text{\tiny{H}}}$}};
		\node at (-3.65,+3.1) {{\small $\phi_{\text{\tiny{H}}}$}};
		\node at (+4.05,+3.1) {{\small $\phi_{\text{\tiny{H}}}$}};
		
		\node at (-3.33,-0.61) {{\scriptsize $\mathcalboondox{V}/M^2$}};
		\node at (-3.33,-1.1) {{\scriptsize $V/M^2$}};
		
		\node at (+4.15,-1.09) {{\scriptsize $\epsilon_{1}$}};
		\node at (+4.15,-1.58) {{\scriptsize $\epsilon_{2}$}};
	\end{tikzpicture}
	\caption{The plots of the Hubble parameter \eqref{Example - special case of HCR solution} (top left panel), number of $e$-folds \eqref{CP Example - Number e-fold HCR} (top right panel), potential \eqref{Example - potential HCR} and effective potential \eqref{Example - effective potential HCR} (bottom left panel), and slow-roll parameters \eqref{Example - e1 for HCR} and \eqref{Example - e2 for HCR} (bottom right panel), once for $\beta_{\text{\tiny{H}}} = -0.5$ and $\gamma=1$ (solid lines) and once for $\beta_{\text{\tiny{H}}} = 0.5$ and $\gamma=-1$ (dashed lines). The parameter $\mathcalboondox{V}_0$ in effective potential is taken to be $1.5$. Vertical grid lines in all panels correspond to the extrema of $H_{,\phi}$ at $\phi_{\text{\tiny{H}}} = d_4 \approx -3.4$.}
	\label{fig: F = gamma x phi HCR}
\end{figure}

We now discuss the HCR solution \eqref{Example CP - solution for HCR conditions}. We consider a special case with $C_4 = 0$ as
\begin{equation}
	H(\phi_{\text{\tiny{H}}}) = M \cosh \left(\phi_{\text{\tiny{H}}}^{3/2} \right), \label{Example - special case of HCR solution}
\end{equation}
where for convenience we have defined $M \equiv C_3$, $\bar{\beta}_{\text{\tiny{H}}}\equiv \sqrt[3]{-2 \gamma \beta_{\text{\tiny{H}}} / 9}$, and $\phi_{\text{\tiny{H}}} \equiv \bar{\beta}_{\text{\tiny{H}}} \phi$. For negative and positive $\beta_{\text{\tiny{H}}}$ the Hubble parameter \eqref{Example - special case of HCR solution} corresponds to the \eqref{Canonical CR - hubble parameter 1} and \eqref{Canonical CR - hubble parameter 2} cases, respectively. We have plotted this parameter for $\beta_{\text{\tiny{H}}} = -0.5$ with $\gamma = 1$, as well as  $\beta_{\text{\tiny{H}}} = 0.5$ with $\gamma = -1$ in Figure \ref{fig: F = gamma x phi HCR} (top left panel). Similar to the FCR case, these two scenarios share a common Hubble parameter but have different dynamical trends. In these specific choices, the parameter $\bar{\beta}_{\text{\tiny{H}}}$ is real and positive. This implies that for negative values of $\phi_{\text{\tiny{H}}}$, the Hubble parameter \eqref{Example - special case of HCR solution} adopts a sinusoidal form, while for positive $\phi_{\text{\tiny{H}}}$ values, it takes on a hyperbolic form. To pinpoint the critical points of the model, we introduce a new parameter $d_n$, defined as $d_n = - (n \pi /2)^{2/3}$. With this definition, zeros of Hubble parameter \eqref{Example - special case of HCR solution} are located at $\phi_{\text{\tiny{H}}} = d_{2s+1}$ and zeros of its derivative, $H_{,\phi} \propto \sqrt{\phi_{\text{\tiny{H}}}} \sinh ( \phi_{\text{\tiny{H}}}^{3/2})$, are located at $\phi_{\text{\tiny{H}}} = d_{2s}$, where $s$ is a non-negative integer. This section focuses on the model's dynamics around 
the first peak of the Hubble parameter on the negative $\phi_{\text{\tiny{H}}}$ axis at $\phi_{\text{\tiny{H}}} = d_4$, and we restrict our discussion to the $d_5<\phi_{\text{\tiny{H}}}<d_3$ interval where Hubble parameter is positive for $M>0$.

Using relation \eqref{number of efolds-HCR}, for the number of $e$-folds we find
\begin{equation}
	N (\phi_{\text{\tiny{H}}}) = \frac{1}{\beta_{\text{\tiny{H}}}} \ln \left( \left| \sinh \left( \phi_{\text{\tiny{H}}}^{3/2} \right) \right|\right) + N_0. \label{CP Example - Number e-fold HCR}
\end{equation}
This function is shown in Figure \ref{fig: F = gamma x phi HCR} (top right panel).
Using \eqref{General CR1 - equation for V} for the potential we get
\begin{equation}
	V = \frac{1}{2} M^2 \left[(\beta_{\text{\tiny{H}}} +3) \cosh \left(2 \phi_{\text{\tiny{H}}}^{3/2} \right)-\beta_{\text{\tiny{H}}} +3\right]. \label{Example - potential HCR}
\end{equation}
For the derivative of effective potential equation \eqref{effective potential-PCR} gives
\begin{equation}
	\mathcalboondox{V}_{,\phi} = \frac{M^2 \bar{\beta}_{\text{\tiny{H}}}^2}{2 \gamma  \phi_{\text{\tiny{H}}} ^2} \left[3 (\beta_{\text{\tiny{H}}} +3) \phi_{\text{\tiny{H}}} ^{3/2} \sinh \left(2 \phi_{\text{\tiny{H}}} ^{3/2}\right)-2 \beta_{\text{\tiny{H}}}  \sinh ^2\left(\phi_{\text{\tiny{H}}} ^{3/2}\right)\right].
\end{equation}
Integrating this equation, the effective potential is derived as
\begin{equation}
	\mathcalboondox{V}= \frac{M^2 \bar{\beta}_{\text{\tiny{H}}}}{2 \gamma  \phi_{\text{\tiny{H}}} } \left[3 \phi_{\text{\tiny{H}}}^{3/2} \left(E_{\frac{2}{3}}\left(2 \phi_{\text{\tiny{H}}}^{3/2} \right)-E_{\frac{2}{3}}\left(-2 \phi_{\text{\tiny{H}}}^{3/2}\right)\right) + 2 \beta_{\text{\tiny{H}}}  \sinh ^2\left(\phi_{\text{\tiny{H}}}^{3/2}\right)\right] + \mathcalboondox{V}_0, \label{Example - effective potential HCR}
\end{equation}
where $\mathcalboondox{V}_0$ is the constant of integration and $E_n(z)$ is the exponential integral which is defined by
\begin{equation}
	E_n(z) = \int_1^{\infty} \frac{e^{-z t}}{t^n} d t.
\end{equation}
Potential \eqref{Example - potential HCR} and effective potential \eqref{Example - effective potential HCR} are plotted in the bottom left panel of Figure \ref{fig: F = gamma x phi HCR}. Zeros of the Hubble parameter and its derivative coincide with extrema of effective potential (see equation \eqref{effective potential-PCR}). Comparing the behavior of effective potential in Figures \ref{fig: F = gamma x phi} and \ref{fig: F = gamma x phi HCR} around the extrema of the Hubble parameter ($a_1$ point in the FCR case and $d_4$ point in the HCR case), we observe that the concavity of both models are alike. This is because, as we mentioned, at this critical point the derivatives of the effective potential in the FCR and PCR conditions are approximately equal, and the concavity of the effective potential is solely determined by the constant-roll parameter, regardless of the parameter $\mathcal{F}$. The concavity of the potential at $d_4$ point, however, depends on the sign of $\mathcal{F}$, which is why in both cases potential is concave.

\begin{figure}
	\centering
	\begin{tikzpicture}
		\node[align=left] (img) at (0,0)
		{\quad\,\,\includegraphics{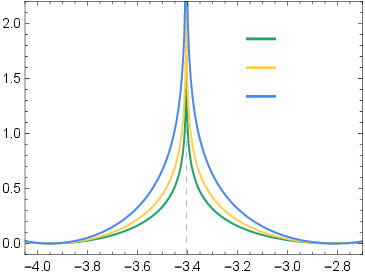}\quad\quad\quad\quad\,\includegraphics{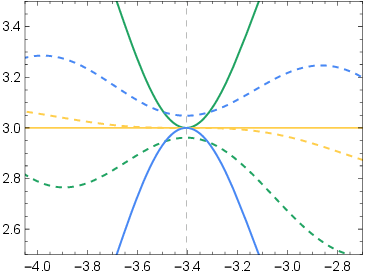}\quad}; //
		
		\node at (-7.25,0.2) {{\small \rotatebox{90}{$N-N_0$}}};
		
		\node at (-3.65,-2.6) {{\small $\phi_{\text{\tiny{H}}}$}};
		\node at (+4.27,-2.6) {{\small $\phi_{\text{\tiny{H}}}$}};
		
		\node at (-1.72,1.60) {{\tiny $\beta_{\text{\scalebox{.8}{H}}} = -4$}};
		\node at (-1.72,1.13) {{\tiny $\beta_{\text{\scalebox{.8}{H}}} = -3$}};
		\node at (-1.72,0.64) {{\tiny $\beta_{\text{\scalebox{.8}{H}}} = -2$}};
		
	\end{tikzpicture}
	\caption{Another plot of the number of $e$-folds \eqref{CP Example - Number e-fold HCR} (left panel), potential \eqref{Example - potential HCR} (solid lines in right panel), and effective potential \eqref{Example - effective potential HCR} (dashed lines in right panel) for $\gamma = 1$ and $\mathcalboondox{V}_0 = 3.382$. Vertical grid line at $\phi_{\text{\tiny{H}}}=d_4 \approx -3.4$ corresponds to the singularity of function $N(\phi_{\text{\tiny{H}}})$.}
	\label{fig: F = gamma x phi 2 HCR}
\end{figure}
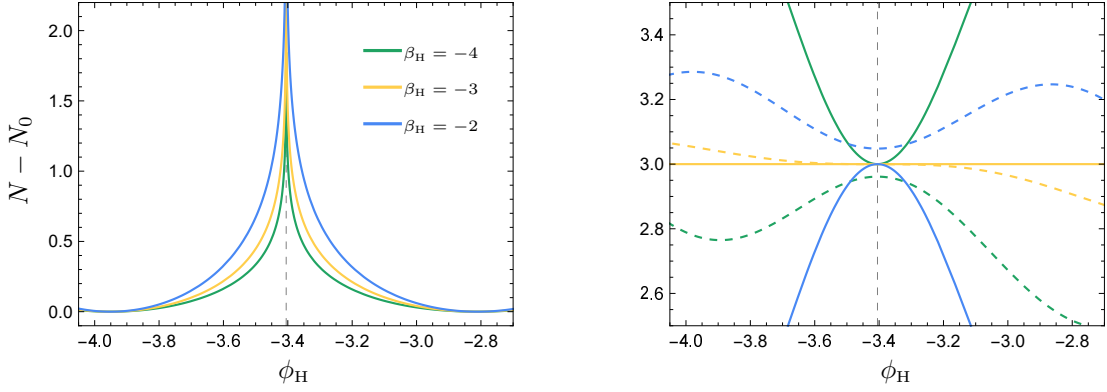

\begin{figure}
	\centering
	\begin{tikzpicture}
		\node[align=left] (img) at (0,0)
		{\quad\,\,\includegraphics{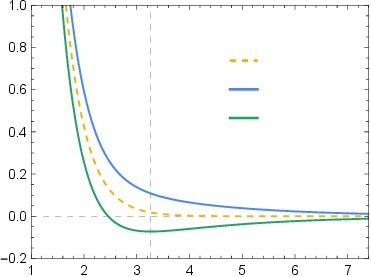}\quad\quad\quad\quad\,\includegraphics{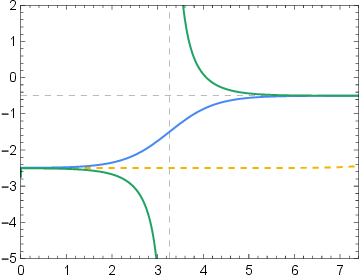}\quad}; //
		
		\node at (-7.25,0.2) {{\small \rotatebox{90}{$(\phi-d_4/\bar{\beta}_{\text{\tiny{H}}}) \times 10^2$}}};
		\node at (+0.57,0.2) {{\small \rotatebox{90}{$\beta_{\text{\tiny{H}}}$}}};
		
		\node at (-3.65,-2.6) {{\small $t$}};
		\node at (+4.27,-2.6) {{\small $t$}};
		
		\node at (-2.16,1.31) {{\scriptsize $\delta = 0$}};
		\node at (-1.86,0.84) {{\scriptsize $\delta = -0.01$}};
		\node at (-1.86,0.35) {{\scriptsize $\delta = +0.01$}};
		
	\end{tikzpicture}
	\caption{We have numerically integrated the field equation \eqref{General CR - equation of motion for Cs constant 3} for $\beta_{\text{\tiny{H}}}=-2.5$, $\gamma=1$, and $M=1$, to evaluate the time evolution of the field (left panel) and the constant-roll parameter $\beta_{\text{\tiny{H}}} = \ddot{H} / (2 H \dot{H})$ (right panel), in order to assess the stability of the constant-roll and propagation model \eqref{Example - special case of HCR solution} under small perturbations of field velocity, $\delta$. Initial values for the field and its velocity are taken as $-2$ and $-1.358-\delta$, respectively. Vertical grid lines correspond to the time when field stops and reverses its direction in the $\delta = +0.01$ case. The horizontal grid line in the right panel shows the corresponding dual attractor solution $\beta_{\text{\tiny{H}}} = -0.5$.}
	\label{fig: stabiliuy analysis - HCR}
\end{figure}

To derive the slow-roll parameters as a function of field, one can use equation \eqref{CP PCR - general e1 behavior} to obtain
\begin{align}
	\epsilon_{1} (\phi_{\text{\tiny{H}}})&= \beta_{\text{\tiny{H}}}  \tan ^2\left(\phi_{\text{\tiny{H}}} ^{3/2}\right), \label{Example - e1 for HCR}\\
	\epsilon_{2} (\phi_{\text{\tiny{H}}})&= 2 \beta_{\text{\tiny{H}}}  \sec^2 \left(\phi_{\text{\tiny{H}}} ^{3/2}\right).\label{Example - e2 for HCR}
\end{align}
The bottom right panel of Figure \ref{fig: F = gamma x phi HCR} represents these parameters. As we can see, at critical point $\phi_{\text{\tiny{H}}} = d_4$, the model exhibits the quasi-canonical behavior with $\epsilon_{1}$ approaching zero and $\epsilon_{2} \approx 2 \beta_{\text{\tiny{H}}}$. One can think of different inflationary scenarios depending on the initial value of the field. In $\beta_{\text{\tiny{H}}} < 0$ case, the field generally moves toward the extrema of the Hubble parameter (corresponding to singularities of $N(\phi_{\text{\tiny{H}}})$) at $d_4$ point, where we observe the quasi-canonical behavior, while in $\beta_{\text{\tiny{H}}} > 0$ case, the field moves away from these points, gradually leaving the quasi-canonical phase and exits the inflation when $\epsilon_{1}$ reaches one. The exception is when both $\beta_{\text{\tiny{H}}}$ and initial value of the field are positive. In this case, field moves toward the positive infinity.


To analyse the concavity of the potential and effective potential near $\beta_{\text{\tiny{H}}} = -3$, we have also plotted these parameters for $\beta_{\text{\tiny{H}}} = -4$, $-3$, and $-2$, in the vicinity of the critical point $\phi_{\text{\tiny{H}}} = d_4$ in Figure \ref{fig: F = gamma x phi 2 HCR}. A comparison of Figures \ref{fig: F = gamma x phi 2} and \ref{fig: F = gamma x phi 2 HCR} reveals that despite the general dissimilarity of the potential and effective potential in the HCR case from the FCR, their concavity mirrors that of the FCR case. We also see that, unlike the FCR case, where the potential does not support the USR profile, here we have a flat potential for $\beta_{\text{\tiny{H}}} = -3$, highlighting the effectiveness of the PCR condition for exploring the USR limit of a model.

To check the stability of the model, we have numerically evaluated and plotted the dynamics of the field and the constant-roll parameter $\beta_{\text{\tiny{H}}}$ in response to slight perturbations in the field velocity in Figure \ref{fig: stabiliuy analysis - HCR}. Similarly to the FCR case, we find that in the unperturbed case the constant-roll parameter $\beta_{\text{\tiny{H}}}$ retains a constant value $-2.5$ throughout the evolution, and upon imposing a tiny perturbation to the initial conditions, the model swiftly transitions into the corresponding dual attractor solution with $\beta_{\text{\tiny{H}}} = -0.5$, and the field eventually settles at the extremum of the effective potential at $\phi_{\text{\tiny{H}}} = d_4$. Despite their shared features, it is essential to acknowledge that the FCR and HCR cases are not equivalent, as the transition trajectory is influenced by both the constant-roll condition and the function $\mathcal{F}$. For instance, as mentioned earlier, the value of the field velocity when the field passes the extrema of effective potential in the FCR case is independent of the form of the function $\mathcal{F}$, whereas it depends on $\mathcal{F}$ in the HCR case (see equations \eqref{CP FCR - dotVeff} and \eqref{effective potential-PCR}).

\section{Constant-Roll Kinetically Driven Inflation and Scalar Perturbations} \label{sec: Constant k Inflation}

In this section, we will examine another example of non-canonical constant-roll models, in which sound speed is not constant and $Q_{,\phi}=0$. The equations of motion for the three different conditions discussed in section \ref{sec: Three constant-roll Conditions} will be solved. For a better comparison between dynamics of the model under these conditions, we will introduce and analyse a concrete inflationary scenario consisting of multiple constant-roll stages. We will discuss the background in detail and study the perturbations. In this scenario, there are phases with non-slow-roll evolution able to amplify the size of curvature perturbation to the level required for triggering primordial black holes (PBHs).

\subsection{Background Dynamics}
We consider a special case of the kinetically driven inflation (or $k$ inflation) as \cite{armendariz1999k, chiba2000kinetically, armendariz2001essentials, chiba2002tracking, malquarti2003new, rendall2006dynamics}
\begin{equation}
	P(X, \phi)=X+\frac{\gamma }{2} X^{2}-V(\phi), \label{Constant k - action}
\end{equation}
where $\gamma$ is a constant with dimension of $M^{-4}$.
Using this relation in equations \eqref{General CR - P(X) equation of motion 1}-\eqref{General CR - P(X) equation of motion 3}, for the Friedmann and field equations we get
\begin{align}
	&6 H^{2}=X \left(2+ 3 \gamma X\right)+ 2 V, \label{Constant k - equations of motion 1} \\
	&\dot{H} + X \left(1+\gamma X \right) = 0, \label{Constant k - equations of motion 2} \\
	&\ddot{\phi} +3 c_s^2 H \dot{\phi} + \mathcalboondox{V}_{,\phi}=0, \label{Constant k - equations of motion 3}
\end{align}
where the sound speed is given by
\begin{equation}
	c_s^2=\frac{1+\gamma X}{1+ 3 \gamma X},
\end{equation}
and the derivative of effective potential \eqref{effective potential-general expression} is related to potential as
\begin{equation}
	\mathcalboondox{V}_{,\phi} = \frac{V_{,\phi}}{1+3 \gamma X}.
\end{equation}
Furthermore, the two first slow-roll parameters are expressed as
\begin{equation}
	\epsilon_{1} = \frac{X (1+ \gamma X)}{H^2}, \quad\quad \epsilon_{2} = 2 \lambda \frac{\ddot{\phi}}{H \dot{\phi}}. \label{constant k - general e2}
\end{equation}
We will also need the time rate of sound speed variation \eqref{Apeendix power - definition of qn}, which is given by
\begin{equation}
	q_1 = \frac{\left(1-c_s^2\right) \left(1-3 c_s^2\right)}{2 c_s^2} \frac{\ddot{\phi}}{H \dot{\phi}}. \label{constant k - general q1}
\end{equation}

We constrain the model using constant-roll conditions \eqref{General CR3 - constant-roll condition 3}, \eqref{General CR1 - fist kind Contant roll condition}, and \eqref{General CR3: constant-roll condition for P(X)}. Using the relation $Q=P+V=X+\frac{\gamma}{2}X^2$, we have
\begin{align}
	\text{FCR:}\,\,&\frac{\ddot{\phi}}{H \dot{\phi}}=\beta_{\text{\tiny{F}}}, \label{Constant k - CR3} \\
	\text{HCR:}\,\,& \lambda \frac{\ddot{\phi}}{H \dot{\phi}}= \beta_{\text{\scalebox{.8}{H}}}, \label{Constant k - CR1} \\
	\text{PCR:}\,\,&\frac{1}{c_s^2}\frac{\ddot{\phi}}{H \dot{\phi}}= \beta_{\text{\tiny{V}}}, \label{Constant k - CR2}
\end{align}
where we have used equation \eqref{Constant k - equations of motion 2} to eliminate $\dot{H}$ and $\ddot{H}$ from equations. Note that here $\lambda$ and $c_s$ are evolving variables and these constant-roll conditions are not homologous in general. From this system of equations, one can see that one important parameter for the evolution is the product $\gamma X$ (or $c_s^2$). For $\gamma X \gg 1$, we have $P \approx \frac{\gamma}{2} X^2- V$, and the kinetic energy is dominated by the non-canonical term, but for $\gamma X\ll 1$, action reduces to the canonical case. In the latter case, the three constant-roll conditions are homologous with $\beta_{\text{\tiny{F}}} = \beta_{\text{\tiny{H}}} = \beta_{\text{\tiny{V}}}$, and the background parameters $\epsilon_n$, $c_s$, and $q_n$ reduce to the canonical values discussed in section \ref{sec: Canonical CR Model}. For $\gamma X \gg 1$, the sound speed is approximated by $c_s^2 \approx 1/3$ with $q_1=0$, which means that no matter which of the constant-roll conditions are used, inflation is in a constant propagation regime with $\mathcal{F}=\gamma / 2$ (see \eqref{General CR - Lagrangian for Cs=constant}), and therefore we can get use of the results discussed in section \ref{sec: constant-roll and Propagation Inflation}. In this regime, since $\mathcal{F}$ is constant, the three models are homologous and the constant-roll parameters are related by $ 6 \beta_{\text{\tiny{F}}} = 3 \beta_{\text{\tiny{H}}} = 2 \beta_{\text{\tiny{V}}}$.

The interesting feature of the model is that it can naturally transition between the two limiting phases. In the case when constant-roll parameter is negative, one can think of an inflationary scenario that starts in $\gamma X \gg 1$ regime and gradually transitions to a canonical constant-roll system. On the contrary, when constant-roll parameter is positive and $X$ is growing in time, the model evolves in the opposite direction, transitioning from the canonical to $\gamma X \gg 1$ phase. Section \ref{sec: k inflation - PBH formation} will present an example where we manipulate the constant-roll parameter to illustrate these transition intervals.

While the three constant-roll conditions are homologous during the two limiting phases, the transition between these phases is strongly influenced by the chosen condition. Here we briefly discuss when the constant propagation approximation breaks and we can expect different consequences from different constant-roll constraints. If the constant-roll parameter is negative (cases \eqref{Canonical CR - hubble parameter 1} and \eqref{Canonical CR - hubble parameter 3}) and we start the inflation in $c_s^2 = 1/3$ phase, we can write
\begin{align}
	\gamma \dot{\phi}^{2}= 2M \sqrt{ \gamma \, |\beta_{\text{\tiny{H}}}|} \, e^{\beta_{\text{\tiny{H}}} \left(N-N_{\text {ini}}\right)}, \label{large gamma - comparing g phi2 with 1}
\end{align}
where we have assumed that inflation starts at $N= N_{\text {ini}}$. Note that this relation is consistent with the general relation \eqref{large gamma - general relation for dotphi} since we have $\beta_{\text{\tiny{H}}} = 2 \beta_{\text{\tiny{F}}}$ in $\gamma \dot{\phi}^{2} \gg 1$ limit. We now define parameter $\delta$ such that the approximation $P \approx \frac{\gamma}{2} X^2 - V$ starts to break at the critical value $\gamma \dot{\phi}_{\text {cri}}^{2} = \delta$.\footnote{It is important to note that the transition between these two phases occurs smoothly. Defining a critical time is merely to demarcate the boundaries of each phase and does not mean a sudden shift from one phase to the other within the model. The critical time serves as a convenient marker to outline the range where our constant propagation approximation becomes less accurate and to highlight the regimes at which constant-roll conditions result in novel consequences and are of particular interest.}
Using equation \eqref{large gamma - comparing g phi2 with 1}, one can find the critical $e$-folds number as
\begin{equation}
	N_{\text {cri}} = \frac{1}{2 \beta_{\text{\tiny{H}}}} \ln \left(\frac{\delta ^2}{4 \gamma M^2 |\beta_{\text{\tiny{H}}}|}\right) + N_{\text {ini}}. \label{large gamma - critical N for negative alpha}
\end{equation}
As an example, if $\delta = 10, M=1, \gamma = 10^5$, and $\beta_{\text{\tiny{H}}}=-0.5$ then $N_{\text {cri}} \approx 7.6 + N_{\text {ini}}$. Therefore, approximation $c_s^2 \approx 1/3 $ is valid for the first 7.6 $e$-folds of the inflation. After this point the two terms in relation \eqref{Constant k - action} are comparable and we must use exact solutions. When $\dot{\phi}$ continues to decay, the model will eventually reach the canonical limit.


Now consider the case when constant-roll parameter is positive (case \eqref{Canonical CR - hubble parameter 2}) and model starts in the canonical phase. To evaluate the moment when canonical approximation starts to break, we use solution \eqref{Canonical CR - Solution 21} to write
\begin{equation}
	\gamma \dot{\phi}^{2}= 2 \gamma M^2 \beta_{\text{\tiny{H}}} e^{2 \beta_{\text{\tiny{H}}} (N-N_{\text {ini}})},
\end{equation}
where we have assumed that inflation starts at $N=N_{\text {ini}}$. Again, this condition is consistent with relation \eqref{large gamma - general relation for dotphi} since in the canonical limit we have $\beta_{\text{\tiny{H}}}=\beta_{\text{\tiny{F}}}$. Similar to \eqref{large gamma - critical N for negative alpha}, by defining a proper $\delta$ parameter, we can find the number of $e$-folds that canonical constant-roll lasts by
\begin{equation}
	N_{\text {cri}} =\frac{1}{2 \beta_{\text{\tiny{H}}}} \ln \left(\frac{\delta }{2 \gamma M^2 \beta_{\text{\tiny{H}}}}\right) + N_{\text {ini}}.
\end{equation} 
For instance, for $\delta = 0.1, M=1, \gamma = 10^{-5}$, and $\beta_{\text{\tiny{H}}}=0.5$ we find $N_{\text {cri}} \approx 9.2 + N_{\text {ini}}$. That means canonical approximation is valid at the beginning of inflation until it breaks at around $9.2$ $e$-folds. After this point, we must use exact solutions. As $\dot{\phi}$ grows, we gradually enter the second constant propagation phase with $c_s^2 = 1/3$.


Let us focus more on the background dynamics of this model. The time derivatives in equations \eqref{Constant k - equations of motion 1} and \eqref{Constant k - equations of motion 2} can be written in terms of $N$ to give
\begin{align}
	3 H^{2} & =\frac{H^{2} \phi_{, N}^{2}}{2}\left(1+\frac{3 \gamma}{4} H^{2} \phi_{, N}^{2}\right)+V, \label{Exact solutions - equation of motion 1} \\
	H_{, N} & =-\frac{H \phi_{, N}^{2}}{2}\left(1+\frac{\gamma}{2} H^{2} \phi_{, N}^{2}\right). \label{Exact solutions - equation of motion 2}
\end{align}
In HCR models, the Hubble parameter is the same as the constant-roll canonical case. For FCR models, however, after writing the time derivatives in constant-roll condition \eqref{Constant k - CR3} in terms of $N$ and removing $\phi_{,N}$ and $\phi_{,NN}$, using equation \eqref{Exact solutions - equation of motion 2}, we get
\begin{align}
	&\gamma H_{,N}^4+H \left(-8 \gamma \beta_{\text{\tiny{F}}} H_{,N}^3+2 \gamma H_{,N}^2 H_{,NN}+2 \beta_{\text{\tiny{F}}} H_{,NN}-4 \beta_{\text{\tiny{F}}}^2 H_{,N}\right) \\
	&+\gamma H^2 \left(H_{,NN}-4 \beta_{\text{\tiny{F}}} H_{,N}\right)^2+2 \beta_{\text{\tiny{F}}} H_{,N}^2=0. \nonumber
\end{align}
Fortunately, this can be integrated to give
\begin{align}
	H^2 = \frac{1}{2 \gamma \beta_{\text{\tiny{F}}}} \left[C_1-\left(C_2 e^{2 \beta_{\text{\tiny{F}}} N}+1\right)^2\right], \label{Exact solutions - Hubble for HCR}
\end{align}
where $C_1$ and $C_2$ are two complex constants, which similar to our discussion in section \ref{sec: Canonical CR Model}, they should be chosen properly so that \eqref{Exact solutions - Hubble for HCR} leads to a viable inflationary scenario.

In order to compare the FCR and HCR conditions, one can proceed to derive other parameters of the model analytically, though the PCR condition can only be treated numerically. To explore these models and show the details of transitions, we would rather perform a full numerical study on an inflationary scenario consisting of multiple constant-roll stages instead. In the following subsection, we construct a four-stage constant-roll model and study the evolution of scalar perturbations in this scenario for the three constant-roll constraints. The time dependent background parameters in each case can affect the horizon crossing time of modes, and besides other mechanisms, translate into a scale dependent amplification of the power spectrum.

\subsection{Evolution of Scalar Perturbations} \label{sec: k inflation - PBH formation}
Having examined the general dynamical properties of the background parameters, we now proceed to investigate scalar perturbations of the model.
The curvature perturbations of wavenumber $k$ for the general single field action \eqref{Appendix - action for PX theories} satisfy the following differential equation:
\begin{equation}
	\mathcal{R}_{k}^{\prime \prime}+2 \frac{z^{\prime}}{z} \mathcal{R}_{k}^{\prime} + c_s^2 k^{2} \mathcal{R}_{k}=0, \label{PBH formation - equation of motion for u}
\end{equation}
where $z \equiv a \sqrt{2 \epsilon_{1}}/c_{s}$ and $z^{\prime} / z$ can be expressed in terms of the background parameters as
\begin{equation}
	\frac{z^{\prime}}{z}=a H \left(1+\frac{\epsilon_{2}}{2}-q_1 \right). \label{PBH formation - general relation for zpoz}
\end{equation}
A full analysis of equation \eqref{PBH formation - equation of motion for u} shows the phenomenological consequences of a departure from slow-roll and/or canonical models. One can (plot the bracket in equation \eqref{PBH formation - general relation for zpoz} to) find the ranges of constant-roll parameter at which $z^\prime / z$ is negative implying that the friction term (proportional to $z^\prime/z$) acts as driving term and the would be decaying solution of \eqref{PBH formation - general relation for zpoz} grows in time. This will contaminate the nearly constant solution $\mathcal{R}_k$, leading to an amplified value at late time, for superhorizon scales. It is clarified that such an (amplitude) enhancement can be responsible for the production of PBHs \cite{ozsoy2023inflation}.

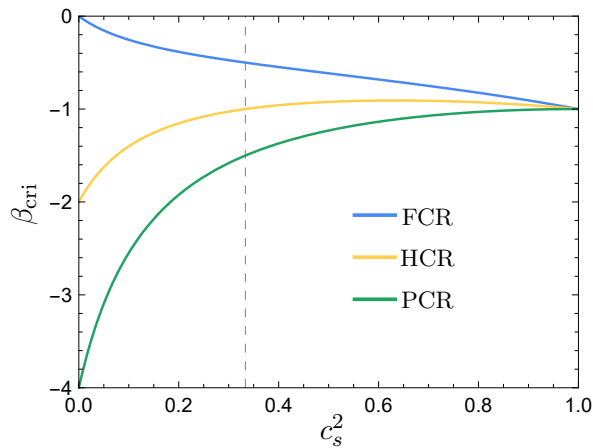
\begin{figure}
	\centering
	\begin{tikzpicture}
		\node[align=left] (img) at (0,0)
		{\includegraphics[scale=1.15]{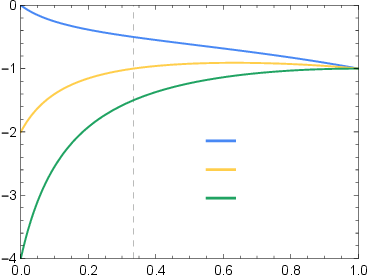}}; //
		
		\node at (-3.9,0.1) {{\rotatebox{90}{$\beta_{\text{cri}}$}}};
		
		\node at (0.15,-2.9) {{$c_s^2$}};
		
		\node at (1.4,-0.07) {{\footnotesize FCR}};
		\node at (1.4,-0.62) {{\footnotesize HCR}};
		\node at (1.4,-1.17) {{\footnotesize PCR}};
		
	\end{tikzpicture}
	\caption{The plot of the $\beta_{\text{cri}}$ function defined in equation \eqref{general PBH formation formula} as a function of $c_s^2$ for the three constant-roll conditions. Selecting a constant-roll parameter smaller than $\beta_{\text{cri}}$ is the sufficient condition for achieving a negative $z^{\prime} / z$, which can lead to mode amplification and a subsequent burst in the power spectrum. The vertical grid line corresponds to the $c_s^2 = 1/3$.}
	\label{fig: critical beta}
\end{figure}

As alluded to above, in canonical constant-roll models $\epsilon_2\approx 2\beta$, hence $\beta < -1$ is the sufficient condition for the mode amplification. In constant-roll models of our interest, $\epsilon_{2}$ and $q_1$ are proportional to the $\ddot{\phi}/(H\dot{\phi})$ which is constrained by equations \eqref{Constant k - CR3}-\eqref{Constant k - CR2}. To discuss the three models altogether, we define a function as
\begin{equation}
	\Lambda (c_s) \equiv \begin{cases}1, & \text{in FCR} \\ \lambda^{-1}, & \text{in HCR} \\ c_s^2, & \text{in PCR}\end{cases}.
\end{equation}
With this definition, the constant-roll conditions can be represented in a unified form, $\ddot{\phi}/(H\dot{\phi}) = \beta \Lambda$, where $\beta$ can be any of the three constant-roll parameters. Using equations \eqref{constant k - general e2} and \eqref{constant k - general q1} we find the general condition to achieve negative $z^\prime / z$ as
\begin{equation}
	\beta < \beta_{\text{cri}} \equiv \frac{2}{ (3 c_s^2 -5) } \frac{1}{\Lambda}. \label{general PBH formation formula}
\end{equation}
In Figure \ref{fig: critical beta}, we have plotted $\beta_{\text{cri}}$ in terms of $c_s^2$ for each constant-roll condition. In the canonical limit, this condition necessitates $\beta < -1$, while in $\gamma X \gg 1$ limit, it reduces to $\beta < -1/2$, $-1$, and $-3/2$ for FCR, HCR, and PCR conditions, respectively.

\begin{figure}
	\centering
	\begin{tikzpicture}
		\node[align=left] (img) at (0,11.4)
		{\includegraphics{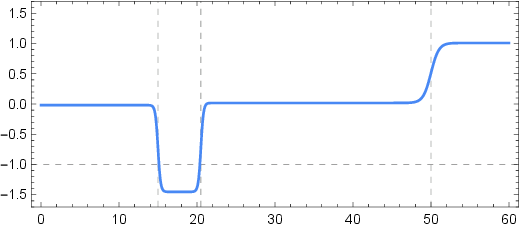}};
		\node[align=left] (img) at (0,5.7)
		{\,\quad\includegraphics{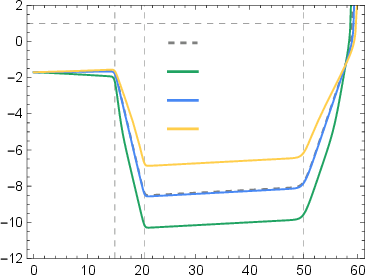}\,\quad\quad\quad\quad\includegraphics{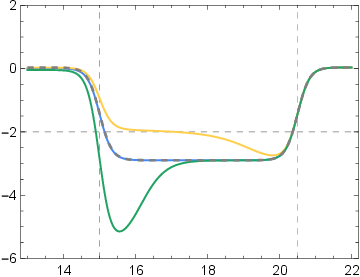}\quad};
		\node[align=left] (img) at (0,0)
		{\,\,\quad\includegraphics{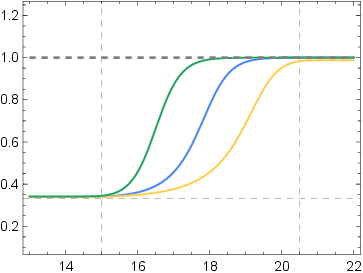}\,\,\quad\quad\quad\,\,\includegraphics{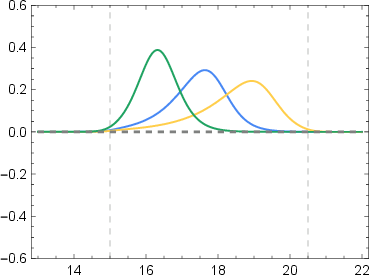}\quad}; //
		
		\node at (-4.55,11.55) {{\small \rotatebox{90}{$\beta$}}};
		\node at (-7.15,5.85) {{\small \rotatebox{90}{$Log(\epsilon_{1})$}}};
		\node at (+0.6,5.85) {{\small \rotatebox{90}{$\epsilon_{2}$}}};
		\node at (-7.15,0.15) {{\small \rotatebox{90}{$c_s^2$}}};
		\node at (+0.6,0.15) {{\small \rotatebox{90}{$q_1$}}};
		
		\node at (-3.65,-2.6) {{\small $N$}};
		\node at (-3.65,+3.1) {{\small $N$}};
		\node at (+0.20,+9.25) {{\small $N$}};
		\node at (+4.05,-2.6) {{\small $N$}};
		\node at (+4.05,+3.1) {{\small $N$}};
		
		\node at (-3.38,10.27) {{(a)}};
		\node at (-5.93,+4.15) {{(b)}};
		\node at (+1.70,+4.15) {{(c)}};
		\node at (-5.93,-1.47) {{(d)}};
		\node at (+1.70,-1.47) {{(e)}};
		
		\node at (-3.08,+7.31) {{\tiny Canonic}};
		\node at (-3.27,+6.82) {{\tiny FCR}};
		\node at (-3.27,+6.34) {{\tiny HCR}};
		\node at (-3.27,+5.85) {{\tiny PCR}};
	\end{tikzpicture}
	\caption{The evolution of background parameters for the multiple-stage constant-roll scenario understudy in section \eqref{sec: k inflation - PBH formation}. Panel: (a) constant-roll parameter $\beta$, (b) first slow-roll parameter $\epsilon_{1}$, (c) second slow-roll parameter $\epsilon_{2}$, (d) sound speed $c_s\label{key}$, and (e) parameter $q_1$. In all panels (except panel (a)) the colour coding is the same. The green, blue, and yellow lines represent FCR, HCR, and PCR conditions, respectively. As is usual in comparative studies, we proposed the $\beta(N)$ function (in panel (a)) as a null hypothesis that makes no difference between models. There are four constant-roll stages with $\beta(N) \approx -0.017$, $-1.45$, $0.018$, and $1$, separated by three vertical grid lines at $N=10$, $20.5$, and $50$ $e$-folds. As shown in this figure, at early times the dynamics are in $\gamma X\gg 1$ regime (with $c_s^2 = 1/3$), but as $X$ decreases, it transitions to a stage in canonical regime. Our comparative study on the constant-roll models described in this section shows that the general behavior is similar but the transition period and its starting time is model dependent. In the PCR model, for example, the field spends more time in the non-canonical regime. The details of the scenario and point-by-point analysis of this figure are given in the text.}
	\label{fig: Constant k - PBH}
\end{figure}


To illustrate a viable model in which a transient non-canonical constant-roll stage can generate the growth in the primordial curvature power spectrum, while satisfying CMB constraints and providing a graceful exit from inflation, we construct a four-stage constant-roll model. We plan to compare the effect of each constant-roll condition on the model; so we define constant-roll parameter as a function of time, $\beta(N)$, and use it in the three constant-roll conditions in a similar fashion. We define the following parameter:\begin{sequation}[\small]
	\beta(N) =-\frac{49}{5} \left(\frac{42}{41}+\tanh (3 (N-15))\right) \left(\frac{39}{40}-\tanh (3 (N-20.5))\right) \left(\frac{28}{27} + \tanh (N-50)\right). \label{PBH formation - construct alpha(N)}
\end{sequation}This parameter (plotted in panel (a) of Figure \ref{fig: Constant k - PBH}) describes four stages of constant-roll inflation (with $\beta= -0.017$, $-1.45$, $0.018$, and $1$, respectively). The stages are connected using proper smooth functions. The constant-roll parameter is set to $-0.017$ in the first stage in order to comply with Planck data (see discussion after equation \eqref{Canonical CR - spectral index 2}). At this stage we normalized the power spectrum to meet the CMB constraints \cite{akrami2020planck}, $\mathcal{P}_s = 2.1 \times 10^{-9}$ at $k_{\text{cmb}} = 0.05 \text{Mpc}^{-1}$. After 15 $e$-folds we switch to a large  $|\beta|$ stage responsible for enhancing the power spectrum. We take the constant-roll parameter smaller than $-1$ to have negative $z^{\prime}/z$, but larger than $-3/2$ in order to preserve the attractor behavior of the model. At $21.5$ $e$-folds, when the enhancement in the spectrum amplitude between small and large CMB scales reaches to $10^6-10^7$, we end this stage. The third stage is another slow-roll stage with $\beta = 0.018$ that is compatible with Planck data (see discussion after equation \eqref{Canonical CR - spectral index}). At the end, we enter a short stage with large positive $\beta$ at 50 $e$-folds, that leads to an increase in $\dot{\phi}$ and accordingly $\epsilon_{1}$, allowing us to reach $\epsilon_{1}=1$ and exit inflation gracefully.

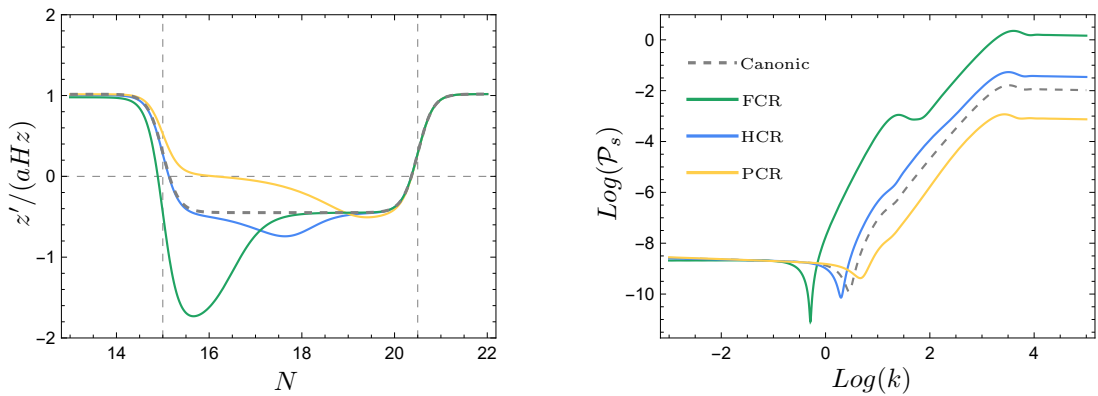
\begin{figure}
	\centering
	\begin{tikzpicture}
		\node[align=left] (img) at (0,0)
		{\,\quad\includegraphics{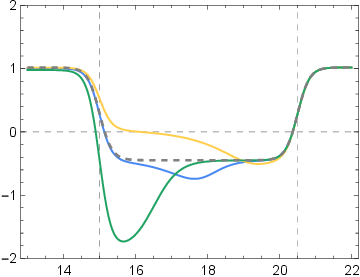}\,\,\quad\quad\quad\quad\includegraphics{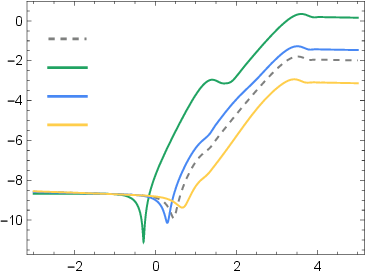}\quad}; //

		\node at (-7.12,0.15) {{\small \rotatebox{90}{$z^{\prime}/(a H z)$}}};
		\node at (+0.6,0.15) {{\small \rotatebox{90}{$Log(\mathcal{P}_s)$}}};
		
		\node at (-3.65,-2.6) {{\small $N$}};
		\node at (+4.05,-2.6) {{\small $Log(k)$}};
		
		\node at (+2.78,+1.60) {{\tiny Canonic}};
		\node at (+2.61,+1.12) {{\tiny FCR}};
		\node at (+2.61,+0.64) {{\tiny HCR}};
		\node at (+2.61,+0.15) {{\tiny PCR}};
	\end{tikzpicture}
	\caption{Parameter $z^{\prime} / (a H z)$ (left panel) and power spectrum $\mathcal{P}_s$ (right panel) for the four-stage constant $k$ scenario described by constant-roll parameter given in \eqref{PBH formation - construct alpha(N)}. The green, blue, and yellow lines represent the evolution constrained by FCR, HCR, and PCR conditions, respectively, and dashed gray curve corresponds to the constant role canonical model.}
	\label{fig: Constant k - PBH 2}
\end{figure}

First, we use parameter \eqref{PBH formation - construct alpha(N)} in equation \eqref{canonic CR - constant-roll condition} and solve the equations of motion and Mukhanov-Sasaki equation \eqref{Appendix - MS equation} numerically to derive power spectrum and dynamical parameters for canonical case. The results are illustrated in Figures \ref{fig: Constant k - PBH} and \ref{fig: Constant k - PBH 2} (dashed gray lines). As can be seen, the power has increased by about $10^7$ times, which is appropriate for PBH formation as suggested in references \cite{ozsoy2023inflation, motohashi2020constant}. In this scenario, as $\beta$ changes, the automatic switch to the corresponding solution will be needed.  The first two stages can be described respectively by solutions \eqref{Canonical CR - hubble parameter 3} and \eqref{Canonical CR - hubble parameter 1}, and solution \eqref{Canonical CR - hubble parameter 2} can be used for the last two stages.

For the non-canonical model \eqref{Constant k - action} the situation is more involved. We repeat the above calculation for  constant-roll conditions \eqref{Constant k - CR3}-\eqref{Constant k - CR2}.  Figures \ref{fig: Constant k - PBH} and \ref{fig: Constant k - PBH 2} show the results (the green, blue, and yellow lines represent FCR, HCR, and PCR conditions, respectively). It should be noted that here we want to compare three constant-roll conditions within a single scenario. We are concerned about the effect of changing the constant-roll condition on the model, not the compatibility of each power spectrum with the data. In this model, we start inflation in $\gamma X \gg 1$ phase. In the first stage, since  $\beta$ is small and $\dot{\phi}$ decays slowly, there is not enough time to transit to the canonical case. During the second stage, $|\beta|$ increases significantly leading to a quick decay in $\dot{\phi}$ and the transition to the canonical case. The transition can be seen even more transparent in $c_s^2$ plot (panel (d) of Figure \ref{fig: Constant k - PBH}). It is evident that the model gradually transitions from a $\gamma X \gg 1$ phase with $c_s^2 = 1/3$ to the canonical phase with $c_s^2 = 1$. During the third stage, $\beta$ becomes positive and $\dot{\phi}$ begins to grow. However, since $\beta$ is small the system stays in the canonical regime throughout. In the last stage, as $\beta$ is large and positive, $\dot{\phi}$ grows dramatically fast and we transit back to $\gamma X \gg 1$ limit before exiting inflation.

Looking at panels (b) and (c) of Figure \ref{fig: Constant k - PBH}, it is evident that the slow-roll parameters for the HCR condition are identical to those for the canonical case. It is due to the fact that for the HCR condition slow-roll parameters can be derived irrespective of action. However, the power spectrum will be different, since the sound speed varies with time. As can be seen in left panel of Figure \ref{fig: Constant k - PBH 2}, the curve $z^{\prime}/(a H z)$ for the HCR model deviates slightly from the canonical case, which is the direct result of varying sound speed (see equations \eqref{PBH formation - general relation for zpoz} and \eqref{Appendix nongaussian - effective potential}).

Panel (b) of Figure \ref{fig: Constant k - PBH} also shows that in all stages of the scenario except the third stage, the slope of $Log(\epsilon_{1})$ differs for three constant-roll conditions. This is because  at third stage we reach the canonical case regardless of our choice of constant-roll condition. This point becomes more clear when we look at the second slow-roll parameter in panel (c) of Figure \ref{fig: Constant k - PBH}. Despite reaching a single path at about 20 $e$-folds, $\epsilon_{2}$ functions are different in earlier times when we have not reached the canonical limit yet. In order to get similar results in early times of inflation when $\gamma X$ is much larger than one we must choose $6\beta_{\text{\tiny{F}}} = 3 \beta_{\text{\tiny{H}}}=2 \beta_{\text{\tiny{V}}}$ (see discussion after equation \eqref{Constant k - CR2}), but in the scenario we utilized $\beta_{\text{\tiny{F}}} =\beta_{\text{\tiny{H}}}= \beta_{\text{\tiny{V}}}=  \beta (N)$, where $\beta (N)$ is defined in \eqref{PBH formation - construct alpha(N)}. However, when we enter the canonical phase, our choice $\beta_{\text{\tiny{F}}} = \beta_{\text{\tiny{H}}}= \beta_{\text{\tiny{V}}}=  \beta (N)$ leads to similar results for three constant-roll conditions.

Looking at the left panel of Figure \ref{fig: Constant k - PBH 2}, we see that in all cases $z^{\prime}/(a H z)$ transiently changes sign, which is necessary to enhance the power spectrum. The more negative it is, the stronger the driving force will enhance the power. Furthermore, we see that the transition time to the canonical phase is different for the three constant-roll conditions; in FCR case the model transitions to canonical case earlier while in PCR case this happens later. That means in FCR case some modes with longer wavelengths (compared to HCR and PCR cases) will also be affected by the enhancement mechanism. Therefore, by looking at $z^{\prime}/(a H z)$ we expect that the FCR condition will result in greater power spectrum enhancement. This can be seen in the right panel of Figure \ref{fig: Constant k - PBH}.

\section{Summary and Conclusion} \label{sec: Summary and Conclusion}
This study explored three methods for imposing the constant-roll condition on non-canonical models. We used a general action as a toy model to introduce a trio of constant-roll conditions, which are all equivalent in the canonical limit but may lead to different results in non-canonical cases. In the first approach, in which a more conventional form of the constant-roll condition in the literature is used, the condition was defined to ensure that the inflaton rolling rate remains constant throughout the inflation. Imposing another type of the constant-roll condition ensures a constant $\epsilon_{2}$ in the model. This second condition enabled us to derive the time profile of the Hubble and slow-roll parameters analytically, even without knowing the exact form of the action. For models with constant sound speed the power spectrum and spectral index of the model were also derived. In these two approaches, we may not obtain a flat USR potential for any value of the constant-roll parameter. In our third approach, we managed to acquire the USR model as constant-roll parameter approaches $-3$. We dubbed these three conditions the FCR, HCR, and PCR conditions, respectively. These conditions were examined to determine under which conditions on action these distinct conditions result in homologous models and when they differ.

As an representative example in our comparative study on the generalized constant-roll models, we investigated a class of non-canonical models characterized by constant sound speed, denoted as "constant propagation inflation". We demonstrated that within this class of models, the HCR and PCR conditions display homologous behavior. In a particular case with $c_s=1$ (corresponding to a class of models known as k/G inflation), we identified two distinct sets of constant-roll solutions under the HCR and FCR conditions. While these solutions generally differ, we showed that in specific limits, where $H_{,\phi}$ approaches zero and the field settles at the extrema of the effective potential, both FCR and HCR models exhibit similar quasi-canonical behavior, where slow-roll parameters mimic their canonical counterparts, though the action generally deviates from the canonical form. We also conducted a full dynamical analysis of the constant-roll parameter under small perturbations in the field velocity to assess the stability of this exemplar of constant propagation model. We determined that for a range of the constant-roll parameters the analytic solutions are unstable and will generally transition to their corresponding dual attractor constant-roll solutions. We noted that the transition trajectory is universal in FCR case but is contingent upon the form of the action in HCR scenario.

We also investigated a kinetically driven non-canonical model with varying sound speed. We observed that when constant-roll parameter is negative, the model transitions automatically from a non-canonical phase to a canonical constant-roll phase, and when constant-roll parameter is positive it evolves in the opposite direction. We noted that while the nature of these two phases remains independent of the chosen constant-roll condition, the transition trajectory is highly affected by this choice. We subsequently introduced  a multi-stage constant-roll scenario and conducted a numerical assessment of the background parameters and the power spectrum of the curvature perturbations. This exploration also demonstrated the implications of model, emphasizing on the sensitivity of the system to the chosen constant-roll condition.

We stress that the constant-roll conditions we defined in this work are not unique. Mathematically, one can define an infinite number of constant-roll conditions for a non-canonical model, each leading to the conventional constant-roll condition $\ddot{\phi} = \beta H \dot{\phi}$ in the canonical limit. We believe this is an intriguing flexibility of the model that has not yet been adequately examined. The study of a non-canonical model under the conventional constant-roll condition can indeed provide some insights. We believe that other forms of the constant-roll condition may also provide valuable information that should not overlooked; so avoiding those in favor of the conventional one is unreasonable.

\section*{Acknowledgements}

We acknowledge the financial support of the research council of the University of Tehran.

\appendix

\section{The Turning Point of \texorpdfstring{$n_s$}{Lg} and \texorpdfstring{$f_{NL}^{re}$}{Lg}} \label{Appendix: Power Specturm}
In this appendix, we go into the mathematical details of the power spectrum and non-Gaussianity in constant-roll scenarios. We specifically explain why in models with constant sound speed the perturbations' behavior changes suddenly at $\beta = -3/2$. We begin by considering the general action \eqref{General CR: general P(X) action},
\begin{equation}
	S=\frac{1}{2} \int d^{4} x \sqrt{-g}[R+2 P(X, \phi)]. \label{Appendix - action for PX theories}
\end{equation}
In this model, the Mukhanov-Sasaki variable $v_{k}$, which is related to the mode function of curvature perturbations $\mathcal{R}_{k}$ through the relation $v_{k} \equiv z \mathcal{R}_{k}$, satisfies the following equation \cite{noller2011non, chen2007observational, seery2005primordial}:
\begin{equation}
	v_{k}^{\prime \prime}+\left(c_{s}^{2} k^{2}-\frac{z^{\prime \prime}}{z}\right) v_{k}=0, \label{Appendix - MS equation}
\end{equation}
where $z \equiv a \sqrt{2 \epsilon_{1}}/c_{s}$ and $c_s$ is the sound speed defined in equation \eqref{General CR - sound speed for P(X)}. In general, $z^{\prime \prime} / z$ term in the above equation can be written as follows:
\begin{equation}
	\frac{z^{\prime \prime}}{z}=a^{2} H^{2}\left(2-\epsilon_{1}+\frac{3}{2} \epsilon_{2}+\frac{1}{4} \epsilon_{2}^{2}-\frac{1}{2} \epsilon_{1} \epsilon_{2}+\frac{1}{2} \epsilon_{2} \epsilon_{3}-3 q_{1}+\epsilon_{1} q_{1}-\epsilon_{2} q_{1}+q_{1}^{2}-q_{1} q_{2}\right), \label{Appendix nongaussian - effective potential}
\end{equation}
where parameters $q_j$ are defined as
\begin{equation}
	q_{1}=\frac{\mathrm{d} \ln c_{s}}{\mathrm{~d} N} , \quad \quad q_{j+1}=\frac{\mathrm{d} \ln q_{j}}{\mathrm{~d} N}. \label{Apeendix power - definition of qn}
\end{equation}
At supper Hubble scales, $k \ll a H$, we can neglect the $c_s^2 k^2$ term in equation \eqref{Appendix - MS equation} and the general solution for corresponding homogeneous curvature perturbation is given by
\begin{equation}
	\mathcal{R}=C_1+C_2 \int \frac{\mathrm{~d} \tau}{z^2}, \label{Appendix power - supper hubble evolution}
\end{equation}
where $C_1$ and $C_2$ are constants of integration. This approximate form is useful for analysing the evolution of modes in large scales and de Sitter space.

We now constrain the model under the HCR condition \eqref{General CR1 - fist kind Contant roll condition},
\begin{equation}
	\frac{\ddot{H}}{2 H \dot{H}}=\beta_{\text{\tiny{H}}}.
\end{equation}
By integrating this equation we find the slow-roll parameters plotted in Figure \ref{fig:Canonic CR - SR parameters}. However, since we are interested in the behavior of perturbations around $\beta_{\text{\tiny{H}}}=-3/2$ point, we will only consider solution \eqref{Canonical CR - hubble parameter 1}. To solve equation \eqref{Appendix - MS equation}, it is also necessary to derive $q_j$ parameters, which cannot be obtained without knowing the form of the action. For simplicity, we consider the constant propagation model \eqref{General CR - Lagrangian for Cs=constant}, where the sound speed is constant. Therefore, $q_j$ parameters are zero, and slow-roll parameters are approximated as $\epsilon_{1} \approx 0$, $\epsilon_{2} \approx 2 \beta_{\text{\tiny{H}}}$, and $\epsilon_{3} \approx 0$ in large positive $N$. Using these results equation \eqref{Appendix - MS equation} simplifies to
\begin{equation}
	v_{k}^{\prime \prime}+\left(c_{s}^{2} k^{2}-\frac{\nu^2 - 1/4}{\tau^{2}}\right) v_{k}=0, \label{Appendix - MS for constant-roll and propagation}
\end{equation}
where we have defined parameter $\nu$ as
\begin{equation}
	\nu \equiv \sqrt{(\beta_{\text{\tiny{H}}}+1)(\beta_{\text{\tiny{H}}}+2) + \frac{1}{4}} = \left|\beta_{\text{\tiny{H}}}+\frac{3}{2}\right| .
\end{equation}
Solving equation \eqref{Appendix - MS for constant-roll and propagation}, we find the standard mode function as follows:
\begin{equation}
	v_{k}(\tau)=\frac{\sqrt{-\pi \tau}}{2} H_{\nu}^{(1)}\left(-c_{s} k \tau\right). \label{Appendix - solution of mode function of perturbations}
\end{equation}
This mode is normalized such that at early times we recover the Bunch-Davis vacuum
\begin{equation}
	\lim _{k \tau \rightarrow - \infty} v_{k}=\frac{1}{\sqrt{2 k}} e^{-i k \tau}.
\end{equation}
We can now evaluate the dimensionless power spectrum of curvature perturbations. With solution \eqref{Appendix - solution of mode function of perturbations}, we will have
\begin{equation}
	\mathcal{P}_s(k) \equiv \frac{k^3}{2 \pi^2}\left|\mathcal{R}_k\right|^2=\frac{H^2}{8 \pi^2 \epsilon_1}\left(\frac{k}{a H}\right)^3 \frac{\pi}{2}\left|H_\nu^{(1)}(-c_s k \tau)\right|^2.
\end{equation}
In $\tau \rightarrow 0$ limit we use the asymptotic formula $\lim _{x \rightarrow 0} H_\nu^{(1)}(x) \simeq-\frac{i}{\pi} \Gamma(\nu)\left(x/2\right)^{-\nu}$ to get
\begin{equation}
	\mathcal{P}_s(k) \approx \frac{H^2}{8 \pi^2 \epsilon_1} \frac{2^{2 \nu-1}[\Gamma(\nu)]^2}{\pi}\left(\frac{c_s k}{a H}\right)^{3-2 \nu},
\end{equation}
which is basically the equation (44) of reference \cite{motohashi2015inflation} with an arbitrary (but constant) sound speed. Accordingly, the spectral index is  $n_s-1 = 3-2\nu = 3- 2 \left|\beta_{\text{\tiny{H}}}+3/2\right|$, which is consistent with equation \eqref{Canonical CR - spectral index}.

Our next step will be the evaluation of the contribution of field redefinition to non-Gaussianity around $\beta_{\text{\tiny{H}}}=-3/2$ in order to investigate its sudden behavior change. In the calculations of non-Gaussianity, it is shown that a field redefinition as $\zeta \rightarrow \zeta_{n} + f(\zeta_{n})$ can significantly simplify the interaction Hamiltonian. In general, for a $P(X,\phi)$ action \eqref{Appendix - action for PX theories}, $f(\zeta)$ is given by
\begin{align}
	f(\zeta)&=\frac{\epsilon_{2}}{4 c_{s}^{2}} \zeta^{2}+\frac{1}{c_{s}^{2} H} \zeta \dot{\zeta}+\frac{1}{4 a^{2} H^{2}}\left[-(\partial \zeta)(\partial \zeta)+\partial^{-2}\left(\partial_{i} \partial_{j}\left(\partial_{i} \zeta \partial_{j} \zeta\right)\right)\right] \label{Appendix - field redefinition F zeta} \\
	&+\frac{1}{2 a^{2} H}\left[(\partial \zeta)(\partial \chi)-\partial^{-2}\left(\partial_{i} \partial_{j}\left(\partial_{i} \zeta \partial_{j} \chi\right)\right)\right].\nonumber
\end{align}
If $\beta_{\text{\tiny{H}}} > -3/2$, modes will freeze out on super Hubble scales. Therefore, all derivative terms in the above equation vanish and do not contribute to the non-Gaussianity.  If $\beta_{\text{\tiny{H}}} < -3/2$, however, $\dot{\zeta}$ term is non-vanishing and will contribute. Therefore, for the calculation of the correlation function we only consider the first two terms in equation \eqref{Appendix - field redefinition F zeta}. After field redefinition $\zeta \rightarrow \zeta_{n}+\frac{\epsilon_{2}}{4 c_{s}^{2}} \zeta_{n}^{2}+\frac{1}{c_{s}^{2} H} \zeta_{n} \dot{\zeta_{n}}$ the three-point function becomes
\begin{align}
	\left\langle\zeta\left(\mathrm{x}_{1}\right) \zeta\left(\mathrm{x}_{2}\right) \zeta\left(\mathrm{x}_{3}\right)\right\rangle&=\left\langle\zeta_{n}\left(\mathrm{x}_{1}\right) \zeta_{n}\left(\mathrm{x}_{2}\right) \zeta_{n}\left(\mathrm{x}_{3}\right)\right\rangle \\
	&+\frac{\epsilon_{2}}{2 c_{s}^{2}}\left(\left\langle\zeta_{n}\left(\mathrm{x}_{1}\right) \zeta_{n}\left(\mathrm{x}_{2}\right)\right\rangle\left\langle\zeta_{n}\left(\mathrm{x}_{1}\right) \zeta_{n}\left(\mathrm{x}_{3}\right)\right\rangle+\mathrm{sym}\right) \nonumber \\
	&+\frac{1}{c_{s}^{2} H}(\langle\dot{\zeta}_{n}\left(\mathrm{x}_{1}\right) \zeta_{n}\left(\mathrm{x}_{2}\right)\rangle\left\langle\zeta_{n}\left(\mathrm{x}_{1}\right) \zeta_{n}\left(\mathrm{x}_{3}\right)\right\rangle+\operatorname{sym})+\mathcal{O}\left(\left\langle\zeta_{n}^{5}(\mathrm{x})\right\rangle\right), \nonumber
\end{align}
where $\mathcal{O}\left(\left\langle\zeta_{n}^{5}(\mathrm{x})\right\rangle\right)$ indicates the higher order terms. Writing the above expression in Fourier space and using relation $\langle{\zeta}_{n}\left(k_{1}\right) \zeta_{n}\left(k_{2}\right)\rangle = (2 \pi)^3 \delta \left(k_1 + k_2\right) \mathcal{R}_{k_1} \mathcal{R}_{k_2}^*$, for the contribution of field redefinition to three-point function we find
\begin{align}
	(2 \pi)^{3} \delta^{(3)}\left(\mathrm{x}_{1}+\mathrm{x}_{2}+\mathrm{x}_{3}\right)\left[\frac{\epsilon_{2}}{2 c_{s}^{2}} \right.&\left.\left(\left|\mathcal{R}\left(\mathrm{x}_{2}\right)\right|^{2}\left|\mathcal{R}\left(\mathrm{x}_{3}\right)\right|^{2}+\mathrm{sym}\right)+\right. \label{Appendix - contribution of field redifiniton in three point function} \\
	\left.\frac{1}{c_{s}^{2} H} \right. & \left. \left(\left|\mathcal{R}\left(\mathrm{x}_{2}\right)\right|^{2} \dot{\mathcal{R}}\left(\mathrm{x}_{3}\right) \mathcal{R}^{*}\left(\mathrm{x}_{3}\right)+\left|\mathcal{R}\left(\mathrm{x}_{3}\right)\right|^{2} \dot{\mathcal{R}}\left(\mathrm{x}_{2}\right) \mathcal{R}^{*}\left(\mathrm{x}_{2}\right)+\mathrm{sym}\right)\right]. \nonumber
\end{align}
We now try to find a relationship between $\dot{\mathcal{R}}$ and $\mathcal{R}$. For this, we use relation $v = z \mathcal{R}$ to write
\begin{equation}
	\dot{\mathcal{R}}=-\mathcal{R} H\left(\frac{\tau v^{\prime}}{v}+\frac{\epsilon_{2}}{2}+1\right). \label{Appendix - write dotu in term of u}
\end{equation}
Using solution \eqref{Appendix - solution of mode function of perturbations} for $\tau v^{\prime} / v$ term in the above equation we can write
\begin{align}
	\lim _{\tau \rightarrow 0} \frac{\tau v^{\prime}}{v} = \lim _{\tau \rightarrow 0} \frac{c_s k \tau H_{\nu +1}^{(1)}(- c_s k \tau)}{H_\nu^{(1)}(- c_s k \tau) } + \nu + \frac{1}{2} \approx -\nu + \frac{1}{2}, \label{Appendix - tau v' / v}
\end{align}
where again we have used asymptotic relation $\lim _{x \rightarrow 0} H_\nu^{(1)}(x) \simeq-\frac{i}{\pi} \Gamma(\nu)\left(x/2\right)^{-\nu}$ for Hankel function. In the light of equations \eqref{Appendix - contribution of field redifiniton in three point function}-\eqref{Appendix - tau v' / v} the bispectrum and non-Gaussianity parameter can be found as
\begin{align}
	&B_{s}^{r e}\left(k_{1}, k_{2}, k_{3}\right)=\frac{1}{c_{s}^{2}}\left(2 \nu -3 - \frac{\epsilon_{2}}{2} \right)\left[\left|u_{k_{2}}\left(\tau_{\mathrm{f}}\right)\right|^{2}\left|u_{k_{3}}\left(\tau_{\mathrm{f}}\right)\right|^{2}+\mathrm{sym}\right], \\
	&f_{N L}^{r e}= \frac{5}{6 c_{s}^{2}}\left( 2 \nu -3 - \frac{\epsilon_{2}}{2} \right) = \frac{5}{6 c_{s}^{2}}\left( 2 \left|\beta_{\text{\tiny{H}}}+\frac{3}{2}\right| -3 - \frac{\epsilon_{2}}{2} \right), \label{Appendix - Fnl}
\end{align}
which for $c_s=1$ and $\epsilon_{2} \approx 2 \beta_{\text{\tiny{H}}}$ leads to relation \eqref{Canonical CR - non-Gaussianity parameter}. Relation \eqref{Appendix - Fnl} was first derived in \cite{namjoo2013violation} for canonical models.

\bibliography{myrefs}{}
\bibliographystyle{JHEP}
	
\end{document}